\documentclass[3p,authoryear]{elsarticle}
\usepackage{epstopdf}
\usepackage{subfigure,graphicx}
\usepackage{float}
\usepackage{amsmath, amsfonts, amssymb}
\usepackage{amsthm}
\usepackage{epsf}
\usepackage{stmaryrd}
\usepackage{amssymb}
\usepackage{leftidx}
\usepackage{color}
\usepackage{mathtools}
\usepackage{placeins}
\usepackage{booktabs}
\usepackage{enumitem}
\usepackage{caption}
\usepackage{multirow}
\usepackage{stackengine}
\usepackage[utf8]{inputenc}
\usepackage[english]{babel}
\usepackage{color}
\usepackage{graphicx}
\usepackage{subcaption}
\usepackage{slashbox}

\theoremstyle{plain}
\newtheorem{theorem}{Theorem}

\newtheorem{remark}[theorem]{Remark}

\def\bfu{{\bf u}}

\def\bfx{{\bf x}}

\def\bfE{{\bf E}}

\def\bfI{{\bf I}}

\def\bfS{{\bf S}}

\def\bfX{{\bf X}}

\def\bfe{{\bf e}}

\def\bfs{{\bf s}}

\def\e0{\varepsilon_0}

\def\s0{\sigma_0}

\long\def\symbolfootnote[#1]#2{\begingroup%
\def\thefootnote{\fnsymbol{footnote}}\footnote[#1]{#2}\endgroup}

\long\def\symbolfootnote[#1]#2{\begingroup%
\def\thefootnote{\fnsymbol{footnote}}\footnote[#1]{#2}\endgroup}

\makeatletter
\renewcommand\@biblabel[1]{}

\makeatother

\begin{document}
\begin{frontmatter}

\title{A Griffith description of fracture for non-monotonic loading \\ with application to fatigue \vspace{0.1cm}}

\author{Subhrangsu Saha}
\ead{saha15@illinois.edu}

\author{John E. Dolbow}
\ead{john.dolbow@duke.edu}

\author{Oscar Lopez-Pamies}
\ead{pamies@illinois.edu}

\address{Department of Civil and Environmental Engineering, University of Illinois, Urbana--Champaign, IL 61801, USA  \vspace{0.05cm}}

\address{Department of Mechanical Engineering and Materials Science, Duke University, Durham, NC 27708, USA \vspace{0.05cm}}

\vspace{-0.3cm}

\begin{abstract}

With the fundamental objective of establishing the universality of the Griffith energy competition to describe the growth of large cracks in solids \emph{not} just under monotonic but under general loading conditions, this paper puts forth a generalization of the classical Griffith energy competition in nominally elastic brittle materials to arbitrary \emph{non-monotonic} quasistatic loading conditions, which include monotonic and cyclic loadings as special cases. Centered around experimental observations, the idea consists in: $i$) viewing the critical energy release rate $\mathcal{G}_c$ \emph{not} as a material constant but rather as a material function of both space $\bfX$ and time $t$, $ii$) one that decreases in value as the loading progresses, this solely within a small region $\Omega_\ell(t)$ around crack fronts, with the characteristic size $\ell$ of such a region being material specific, and $iii$) with the decrease in value of $\mathcal{G}_c$ being dependent on the history of the elastic fields in $\Omega_\ell(t)$. By construction, the proposed Griffith formulation is able to describe any Paris-law behavior of the growth of large cracks in nominally elastic brittle materials for the limiting case when the loading is cyclic. For the opposite limiting case when the loading is monotonic, the formulation reduces to the classical Griffith formulation. Additional properties of the proposed formulation are illustrated via a parametric analysis and direct comparisons with representative fatigue fracture experiments on a ceramic, mortar, and PMMA. 

\keyword{fracture; fatigue; Paris law; energy methods; brittle materials}
\endkeyword

\end{abstract}

\end{frontmatter}

\section{Introduction} \label{Sec: Introduction}

A fundamental problem in mechanics that has remained open for decades is the description of the growth of large cracks in solids that are subjected to arbitrary \emph{non-monotonic} loading conditions, which include monotonic and cyclic loading conditions as the two limiting cases that, by far, have received the most attention. In this paper, we propose to address this fundamental problem by formulating a generalization of the classical Griffith energy competition \citep{Griffith21} to describe the growth of large cracks in nominally elastic brittle materials under general quasistatic loading conditions. The proposed formulation aims thus at establishing the hegemony of the Griffith energy competition as the principle that describes the growth of large cracks in solids, irrespective of how loads are applied. More broadly, the formulation aims at taking a further step in the quest of a universal macroscopic theory of fracture. 

\begin{figure}[H]
\centering
\centering\includegraphics[width=0.80\linewidth]{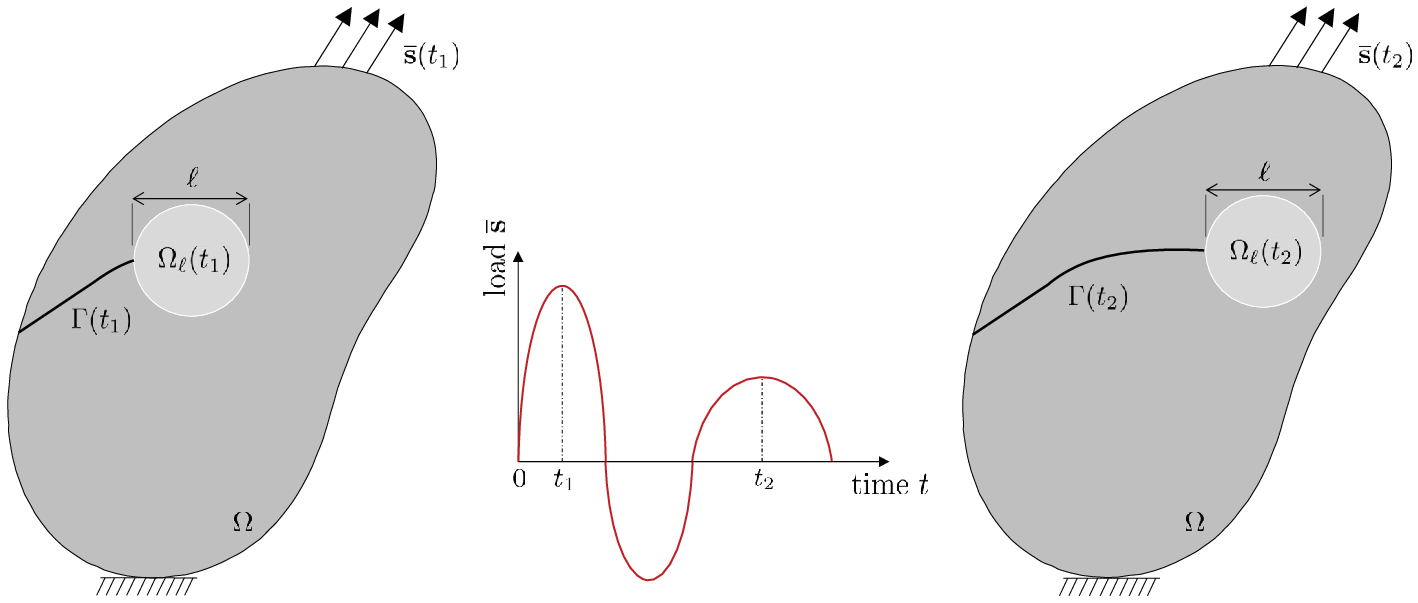}
\caption{\small Schematic of the growth of a large crack $\Gamma(t)$ in a nominally elastic brittle material under non-monotonic loading conditions according to the proposed Griffith formulation. The Griffith criticality condition (\ref{Griffith-criticality}) governs when the large crack grows. In that condition, the critical energy release rate $\mathcal{G}_c$ is \emph{not} a constant but rather a material function of the loading history, as characterized by an internal memory variable $\mathfrak{g}(t)$ of the form (\ref{Int-G}). Specifically, the value of $\mathcal{G}_c$ decreases as the loading progresses. It does so only within a small degradation region $\Omega_{\ell}(t)$, of material-specific characteristic size $\ell$, around the evolving crack front, while it maintains its initial value elsewhere in the body.}\label{Fig1}
\end{figure}

In a nutshell, as schematically depicted in Fig. \ref{Fig1}, the idea behind the proposed formulation --- which, as elaborated below, is centered around a vast body of experimental observations that have been gathered since the 1950s for a variety of nominally elastic brittle materials ranging from rubbers, to ceramics, to rocks --- is threefold:
\begin{itemize}

\item{Much like for monotonic loading, the Griffith criticality condition 
\begin{equation}\label{Griffith-criticality}
-\dfrac{\partial\mathcal{W}}{\partial \Gamma}=\mathcal{G}_c
\end{equation}
is taken to govern when large cracks grow, with the caveat that the critical energy release rate $\mathcal{G}_c$ is \emph{not} a material constant --- as assumed under monotonic loading --- but rather a material function of both space $\bfX$ and time $t$,}

\item{in particular, $\mathcal{G}_c$ is a non-negative material function that decreases in value as the loading progresses, this solely within a small degradation region
\begin{equation*}
\Omega_{\ell}(t)
\end{equation*}
around crack fronts, with the characteristic size $\ell$ of such a region being material specific,
}

\item{and with the decrease in value of $\mathcal{G}_c$ being dependent on the loading history, as described by a suitably defined internal memory variable, of the generic form
\begin{equation}\label{Int-G}
\mathfrak{g}(t)=\displaystyle\int_{0}^{t}\displaystyle\int_{\Omega_{\ell}(t)}\,\mathcal{M}\left(\bfE(\bfu(\bfX,\tau)),\tau\right){\rm d}\bfX\,{\rm d}\tau,
\end{equation}
that keeps track of the history of the strain field $\bfE(\bfu)$ around the evolving crack fronts, from the start at time $t=0$ until the current time $t\in[0,T]$.
}

\end{itemize}
Making use of standard notation, the left-hand side in (\ref{Griffith-criticality}) stands for the change in potential energy $\mathcal{W}$ in the body with respect to an added surface area $\Gamma$ from an existing crack, while $\bfu=\bfu(\bfX,t)$ in (\ref{Int-G}) stands for the displacement field. 

By now, since the pioneering cyclic loading experiments and analyses of \citet{Thomas58} on rubbers, \citet{Paris62} on metals, and \citet{Evans74} on ceramics, there is plenty of experimental evidence that shows that a criticality condition of Griffith type --- that is, a criticality condition based on the energy release rate $-\partial\mathcal{W}/\partial \Gamma$ reaching some critical value --- can be used to describe the growth of large cracks in nominally elastic brittle materials subjected to non-monotonic loading conditions; see, e.g., the reviews \citep{MarsFatemi02,Pinter2023} for rubbers and \citep{Ritchie91,MunzFett99} for ceramics. The proposed Griffith criticality condition (\ref{Griffith-criticality}) is consistent with this evidence and makes it precise that it is the value of a loading-history dependent material function\footnote{The creation of new surface by fracture is typically accompanied by the dissipation of energy by deformation (e.g., viscous and/or plastic deformation) around the crack front, as well as possibly elsewhere in the domain occupied by the body of interest, because of the large deformation rates and the large deformations taking place around the crack front. For nominally elastic brittle materials, these additional sources of energy dissipation are negligible relative to that associated with the creation of new surface itself. In the event that they are not negligible, they would need to be taken into account. Still, in that scenario, the definition of $\mathcal{G}_c$ would remain the same as that given here, as $\mathcal{G}_c$ describes the \emph{intrinsic} energy dissipated by the creation of new surface, irrespective of additional dissipation phenomena that may take place around the crack front; see, e.g., the recent analysis of Griffith fracture in viscoelastic elastomers by \cite{SLP23}.} $\mathcal{G}_c$ --- and \emph{not} just a material constant --- that determines the growth of large cracks.

At present, there is also plenty of experimental evidence that shows that the region around the fronts of large cracks in a nominally elastic brittle material that is mechanically loaded can experience damage before the cracks grow. While this damage can be of different natures depending on the material, it involves invariably the breaking of molecular bonds; see, e.g., \citep{Creton14,Creton20} for rubber, \citep{Ritchie99,Ritchie2013} for ceramics, and also \citep{White14} for more complex materials. This implies that the surface energy density associated with the growth of the cracks --- that is, the value of $\mathcal{G}_c$ around crack fronts --- will be smaller under a non-monotonic loading path, if damage does indeed occur, than under a monotonic one, since the eventual creation of new surface then would involve breaking a smaller number of bonds. The proposed Griffith criticality condition (\ref{Griffith-criticality}), where $\mathcal{G}_c$ decreases over a region $\Omega_{\ell}(t)$ of size $\ell$ around crack fronts as the loading progresses, is consistent with this evidence.

In the sequel, for clarity of exposition, we restrict attention to the basic setting of small-strain kinematics and linear elastic brittle materials; the more general case of finite deformations and nonlinear elastic brittle materials will be considered elsewhere. 

We begin in Sections \ref{Sec: Plate} and \ref{Sec: Griffith description} by presenting the main ideas behind the proposed formulation for the elementary problem of a plate containing a crack in its center that is subjected to a cyclic tensile load. This is a setting that allows for a fully explicit analysis and hence proves particularly instructive. The formulation is then put forth in its general form --- that is, as it applies to large cracks in bodies of arbitrary geometry, made of a linear elastic brittle material, that are subjected to arbitrary non-monotonic quasistatic loading conditions  --- in Section  \ref{Sec: General}. In Section \ref{Sec: Validation}, we provide a first set of validation results by comparing the predictions generated by the proposed formulation with representative fatigue fracture experiments on a silicon nitride ceramic \citep{Ritchie95},  mortar \citep{Wang91}, and PMMA \citep{Clark90}. We close by recording a number of final comments in Section \ref{Sec: Final Comments}.

\subsection{A comment on previous related approaches} 

Before proceeding with the presentation of the proposed Griffith formulation \emph{per se}, it is instructive to provide a summary of related approaches in the literature. There are essentially two types: ($i$) bottom-up and ($ii$) top-down approaches. Notably, these two approaches have been pursued independent of one another. What is more, both of them have been pursued with limited guidance from experimental observations.

\paragraph{Bottom-up approaches} The first theoretical attempt to describe the growth of large cracks in solids subjected to non-monotonic loading conditions by means of a \emph{de facto} evolving critical energy release rate $\mathcal{G}_c$ over a small region around crack fronts appears to be that of \citet{Weertman66}; see also \citet{Weertman73}. The focus of this author was in metals and so he proposed to link the decrease of $\mathcal{G}_c$ to the distribution of dislocations around crack fronts during cyclic loading by making use of a theory that had been previously shown to agree with the classical Griffith criticality condition for the growth of a large crack in a plate under monotonic tension \citep{Bilby63}. Precisely, for the elementary problem of a plate containing a crack of size $2a$ in its center that is subjected to a cyclic tensile stress of constant maximum and minimum values $\sigma_{max}>0$ and $\sigma_{min}=0$, \citet{Weertman66} obtained the Paris law 
\begin{equation*}
\Delta a=C\left(\sigma_{max}\,\sqrt{a}\right)^4
\end{equation*}
for the increment in the half size of the crack per loading cycle, where $C$ is a known constant (not spelled out here). This result happened to be in agreement with a wide range of data available at the time for metals \citep{Paris63}. In spite of this agreement, the approach has not been extended to general crack geometries and boundary conditions, presumably because of the technical difficulties involved.

About four decades later, based on a one-dimensional analysis presented by \cite{Marigo06}, which in turn was motivated by earlier numerical studies \citep{Ravi01,Ortiz01,Geubelle05}, \cite{Marigo2010} proposed to consider a \emph{de facto} evolving critical energy release rate $\mathcal{G}_c$ over a small region around a crack front by making use of a cohesive fracture description of the growth of the crack. In space dimension two, for an isotropic linear elastic material with a cohesive fracture behavior of Dugdale type \citep{Dugdale60}, under several additional assumptions (including certain irreversibility and certain scaling of the length over which cohesive forces are active around the crack front), these authors showed that their framework described the growth of a large crack under monotonic loading according to the classical Griffith criticality condition, at the same time that, under cyclic loading, it described the growth of the crack according to a formula of the form 
\begin{equation*}
\dfrac{{\rm d} a}{{\rm d} N}=C\Delta K_{\texttt{I}}^4,
\end{equation*}
where $a$ is the length of the crack, $N$ is the number of loading cycles, $C$ is a constant, and $\Delta K_{\texttt{I}}$ is the range of stress intensity factors in Mode $\texttt{I}$ resulting from the applied cyclic loading. That is, much like \citet{Weertman66}, \cite{Marigo2010} obtained a Paris law behavior with exponent 4. At present, the reasons for this particular exponent are still unclear. By the same token, it is also not known whether other exponents could be obtained by changing the type of assumed cohesive forces. Addressing these open problems proves crucial to be able to model the growth of cracks in nominally elastic brittle materials at large, since many such materials (e.g., rubber, ceramics, rocks) typically exhibit Paris laws with exponents that differ significantly from 4; see, e.g., the monograph by \citet{Suresh2004}.

In short, while limited in scope and different in their description of the microscopic damage that takes place around crack fronts due to mechanical loading, the works of \citet{Weertman66} and of \cite{Marigo2010} provide insightful examples of bottom-up approaches that illustrate: $i$) that an evolving critical energy release rate $\mathcal{G}_c$ can, in principle, be linked directly to microscopic damage mechanisms localized around crack fronts and $ii$) that such an evolving critical energy release rate $\mathcal{G}_c$ can describe crack growth both under monotonic and cyclic loading conditions. 

\paragraph{Top-down approaches} More recently, within the regularized setting of phase-field models of fracture, top-down approaches have also been proposed in which the critical energy release rate $\mathcal{G}_c$ is \emph{not} kept constant but rather is allowed to evolve in a manner that is \emph{not} derived from microscopic damage mechanisms but instead is prescribed from the outset and fitted to macroscopic data from cyclic loading tests. 

A review of such a type of models --- as well as other types of fatigue phase-field models \citep{Loetal2019,Zohdi20} --- has been recently presented by \cite{Kalinaetal2023}; they include, for instance, the models of \cite{Mesgarnejadetal19,Carraraetal2020,Baxevanis21,Karma22}. Invariably, all these models propose the use of a critical energy release rate $\mathcal{G}_c$ that decreases \emph{pointwise} in space as the loading progresses depending on the history of the stored elastic energy density in the \emph{entirety} of the domain occupied by the body. This dependence, however, appears to be at odds with the very definition of  $\mathcal{G}_c$ --- which physically stands for the energy dissipated per unit area of the crack surface that is created --- and with the experimental observations mentioned above, which, again, indicate that the decrease of $\mathcal{G}_c$ is restricted to a material-specific small region around crack fronts.

Most recently, motivated by the aforementioned works on phase-field models, within the setting of sharp fracture in isotropic linear elastic materials in space dimension two,  \cite{AlessiUlloa2023} have also proposed the prescription of a critical energy release rate $\mathcal{G}_c$ that decreases \emph{pointwise} as the loading progresses depending on the history of the stored elastic energy density. Beyond the pointwise decrease of $\mathcal{G}_c$, another key aspect of their proposed functional dependence is the removal of the singularity of the stored elastic energy density at crack fronts. Having to know what the singular fields are at the crack fronts would appear to make the approach impractical beyond isotropic linear elastic materials in space dimension two.

\section{The elementary problem of a large crack in a plate under cyclic tension}\label{Sec: Plate}

\subsection{The problem}\label{Sec: Plate Problem}

%
\begin{figure}[H]
\centering
\centering\includegraphics[width=0.75\linewidth]{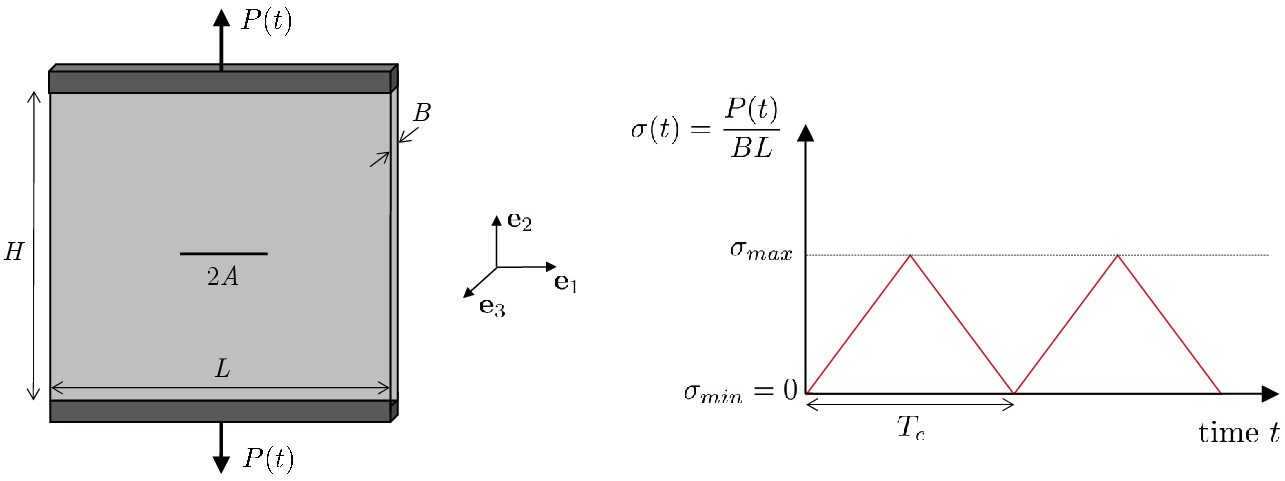}
\caption{\small Schematic of a large thin rectangular plate, containing a large pre-existing crack of initial size $2A$ in its center, that is subjected to a linear cyclic tension, with constant maximum stress $\sigma_{max}>0$, constant minimum stress $\sigma_{min}=0$, and cycle duration $T_c$.}\label{Fig2}
\end{figure}
%

\paragraph{Initial configuration and kinematics} Consider a specimen in the form of a large thin rectangular plate of width $L$ and height $H$ in the $\bfe_1$ and $\bfe_2$ directions and constant thickness $B$ in the $\bfe_3$ direction; see Fig. 2. The specimen contains a large pre-existing central crack of initial length $2A$ --- and hence of initial area $2AB$ --- in the $\bfe_1$ direction, such that $L,H\gg 2A$ and $2A\gg B$; that is, the crack is much smaller than the in-plane dimensions of the plate, but much larger than its thickness. Here, $\{\bfe_1,\bfe_2,\bfe_3\}$ stands for the laboratory frame of reference. We place its origin at the geometric center of the specimen so that, in its initial configuration at time $t=0$, the specimen occupies the domain
\begin{equation*}
\overline{\Omega}=\{\bfX:\bfX\in\mathcal{P}\setminus\Gamma_0\},
\end{equation*}
where
\begin{equation*}
\mathcal{P}=\left\{\bfX: |X_1|\leq\dfrac{L}{2},\,|X_2|\leq\dfrac{H}{2},\,|X_3|\leq\dfrac{B}{2}\right\} \quad \textrm{and} \quad 
\Gamma_0=\left\{\bfX: |X_1|\leq A,\,X_2=0,\,|X_3|\leq\dfrac{B}{2}\right\}.
\end{equation*}
At a later time $t\in(0,T]$, due to the applied boundary conditions described below, the position vector $\bfX$ of a material point in the specimen will move to a new position specified by
\begin{equation*}
\bfx=\bfX+\bfu(\bfX,t),
\end{equation*}
where $\bfu(\bfX,t)$ is the displacement field. We write the associated strain at $\bfX$ and $t$ as
\begin{equation*}
\bfE(\bfu)=\dfrac{1}{2}\left(\nabla\bfu+\nabla\bfu^T\right).
\end{equation*}

\paragraph{Constitutive behavior} The specimen is taken to be made of a homogeneous, isotropic, linear elastic brittle material. Its mechanical behavior is thus characterized by three intrinsic properties: ($i$) its elasticity, ($ii$) its strength, and ($iii$) its critical energy release rate. 

For the problem at hand, the strength does not play a role and hence there is no need to prescribe it here \citep{LP23}. The notion of critical energy release rate $\mathcal{G}_c$ is addressed in Section \ref{Sec: Griffith description} below. As for the elasticity, granted homogeneity and isotropy, the elastic behavior of the material is characterized by the stored-energy function
\begin{equation*}
W(\bfE(\bfu))=\dfrac{E}{2(1+\nu)} {\rm tr}\,\bfE^2+\dfrac{E\nu}{2(1+\nu)(1-2\nu)}\left({\rm tr}\,\bfE\right)^2,
\end{equation*}
where $E$ and $\nu$ are the Young's modulus and the Poisson's ratio, or, by the same token, by the stress–strain relation
\begin{equation*}
\boldsymbol{\sigma}(\bfX,t)=\dfrac{\partial W}{\partial\bfE}(\bfE(\bfu))=\dfrac{E}{1+\nu} \bfE+\dfrac{E\nu}{(1+\nu)(1-2\nu)}\left({\rm tr}\,\bfE\right)\bfI.
\end{equation*}

\paragraph{Applied loading conditions} The specimen is subjected to a cyclic tensile force, of cycle duration $T_c$ and constant amplitude $P_{max}>0$, applied at the top $\partial\Omega^{\mathcal{T}}=\{\bfX:|X_1|< L/2,\,X_2=H/2,\,|X_3|< B/2\}$ and bottom $\partial\Omega^{\mathcal{B}}=\{\bfX:|X_1|< L/2,\,X_2=-H/2,\,|X_3|< B/2\}$ boundaries in the $\bfe_2$ direction, of the linear form
\begin{equation*}
P(t)=
\left\{\begin{array}{ll}
2\left[\dfrac{t}{T_c}-(N-1)\right]P_{max}, & (N-1)T_c\leq t \leq (N-1)T_c+\dfrac{T_c}{2}\vspace{0.2cm}\\
2\left[-\dfrac{t}{T_c}+N\right]P_{max}, & (N-1)T_c+\dfrac{T_c}{2}< t \leq N T_c
\end{array}\right. ,
\end{equation*}
where $N\in \mathbb{Z}^{+}$ is the number of loading cycles. This results in the global cyclic tensile stress
\begin{equation}\label{eq_triangluar_load}
\sigma(t)=\dfrac{P(t)}{B L}=
\left\{\begin{array}{ll}
2\left[\dfrac{t}{T_c}-(N-1)\right]\sigma_{max}, & (N-1)T_c\leq t \leq (N-1)T_c+\dfrac{T_c}{2}\vspace{0.2cm}\\
2\left[-\dfrac{t}{T_c}+N\right]\sigma_{max}, & (N-1)T_c+\dfrac{T_c}{2}< t \leq N T_c
\end{array}\right. ,
\end{equation}
where $\sigma_{max}=P_{max}/(BL)$; see Fig. \ref{Fig2}. 

The cycle duration $T_c$ in (\ref{eq_triangluar_load}) is assumed to be long enough that inertia can be neglected. The value of the maximum stress $\sigma_{max}>0$ is also assumed to be large enough that body forces can be neglected.  

\subsection{The growth of the crack according to experimental observations}

According to numerous experimental observations in a variety of nominally elastic brittle materials, provided that the applied maximum stress $\sigma_{max}$ is large enough to lead to crack growth, but not too large to lead to brutal crack growth, the crack in the specimen  will grow symmetrically in the $\pm\bfe_1$ directions according to a Paris law of the form 
\begin{equation}\label{Paris-Plate}
\dfrac{{\rm d} a}{{\rm d} N}=C\Delta K_{\texttt{I}}^m=C\left(\sigma_{max}\sqrt{\pi a}\right)^m,
\end{equation}
where $a(t)$ stands for the half size of the crack at the current time $t$, $C$ and $m$ are positive constants, and where we recall that $\Delta K_{\texttt{I}}$ denotes the range of stress intensity factors in Mode $\texttt{I}$ resulting from the applied cyclic loading (\ref{eq_triangluar_load}). In the elementary case at hand, granted that the crack remains far away from the boundaries of the plate, $\Delta K_{\texttt{I}}=\sigma_{max}\sqrt{\pi a}$; see, e.g., \citet{Tada73}.

Now, assuming that the growth of the crack is differentiable in time $t=N T_c$ (more on this below), the ordinary differential equation (\ref{Paris-Plate}) implies that the half size of the crack grows according to
\begin{equation}\label{crack_evolution}
a(t)=\left\{\begin{array}{ll}
\left(A^p+\dfrac{q}{T_c}t\right)^{1/p}\quad \textrm{with}\quad
\left\{\begin{array}{l} p=1-\dfrac{m}{2}\vspace{0.2cm}\\
q=\left(1-\dfrac{m}{2}\right)C(\sigma_{max}\sqrt{\pi})^m\end{array}\right. & \quad {\rm if}\quad m\neq 2 \vspace{0.2cm}\\
A\, e^{\frac{C\sigma^2_{max} \pi}{T_c} t} & \quad {\rm if}\quad m= 2\end{array}\right. .
\end{equation}

\begin{remark}
\emph{In practice, the half size $a(t)$ of the crack would be measured every certain number of cycles from which an average ${\rm d} a/{\rm d} N$ would then be computed. Thus, the \emph{direct measurement} would be that of the evolution (\ref{crack_evolution}) of the crack size and \emph{not} that of its rate (\ref{Paris-Plate}).}
\end{remark}

\begin{remark}
\emph{The equation (\ref{crack_evolution}) typically applies up to the instance in time, say $T_{\texttt{m}}$, when the crack becomes unstable under monotonic loading, in the present case, at most up to
\begin{equation*}
T=T_{\texttt{m}}=\left\{\begin{array}{ll}
\dfrac{T_c}{q}\left[\left(\dfrac{G_c E}{\sigma_{max}^2\pi}\right)^p-A^p\right] & \quad {\rm if}\quad m\neq 2 \vspace{0.2cm}\\
\dfrac{T_c}{C \sigma^2_{max}\pi}\ln\left[\dfrac{G_c E}{\sigma_{max}^2\pi A}\right] & \quad {\rm if}\quad m= 2\end{array}\right. .
\end{equation*}
In this last expression, as elaborated in the next section, $G_c$ denotes the \emph{initial} critical energy release rate of the material.
}
\end{remark}

\subsubsection{Crack growth by jumps}

An aspect that is fundamental in understanding how cracks grow under cyclic loadings, but that still remains unresolved --- and that, surprisingly, is seldom discussed in the literature --- is whether, \emph{within the continuum view of matter}, cracks grow via a \emph{continuous} or via a \emph{discontinuous} process (i.e., by jumps) in time. If the former, then equation (\ref{crack_evolution}) would apply for all $t\in[0,T]$. If the latter, then equation (\ref{crack_evolution}) would apply only at time discrete values 
\begin{equation}\label{Discrete tr}
t_j\in\{0=t_0,t_1,...,t_{J-1},t_{J}=T\},
\end{equation}
those at which the crack jumps.

In this work, we shall favor the view that under cyclic loading conditions cracks grow by jumps. More specifically, we consider that under cyclic loading conditions the growth of cracks occurs by jumps of length scale $\ell$, a length scale that is material specific\footnote{Interestingly, growth of cracks by jumps has also been observed in some models for elasto-plastic materials under \emph{monotonic} loading conditions; see, e.g., \cite{DalMasso22}.}. 

\begin{figure}[H]
\centering \includegraphics[scale=0.48]{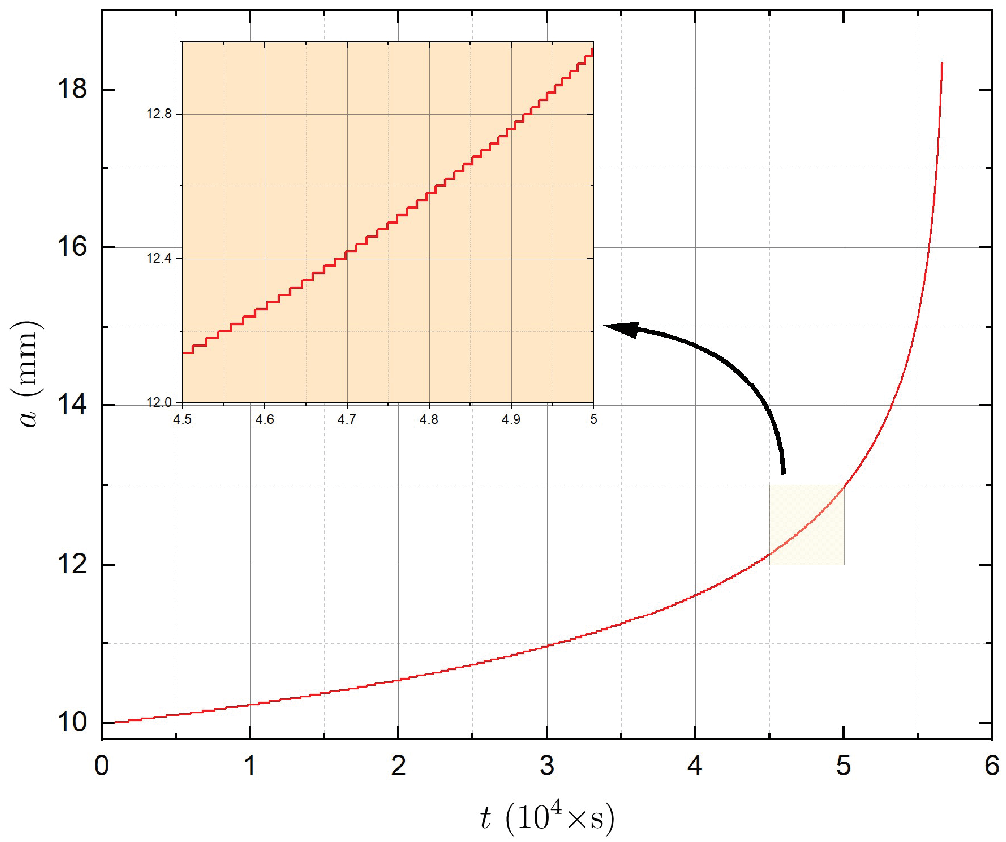}
\caption{Representative time-discontinuous evolution of the half size (\ref{crack_evolution}) of a crack, of initial half size $A=10$ mm, by jumps of  constant size $\ell=20$ $\mu$m. The result pertains to the case of $\sigma_{max}=25$ MPa, $T_c=0.02$ s, $T=T_{\texttt{m}}=5.6615\times 10^4$ s, $E=300$ GPa, $G_c=120$ N/m, $m=18$, and $C=1.01 \times  10^{-21}$ MPa$^{-18}$ m$^{-8}$, which are representative values for silicon nitride \citep{Ritchie91}.}
\label{Fig3}
\end{figure}

By way of an example, Fig. \ref{Fig3} provides the plot of such a time-discontinuous growth process for a crack of initial half size $A=10$ mm for the case of $\sigma_{max}=25$ MPa, $T_c=0.02$ s, $T=T_{\texttt{m}}=5.6615\times 10^4$ s, $E=300$ GPa, $G_c=120$ N/m, $m=18$, $C=1.01 \times  10^{-21}$ MPa$^{-18}$ m$^{-8}$, and $\ell=20$ $\mu$m, which are values representative of the ceramic silicon nitride \citep{Ritchie91}. For this example, since $m\neq2$, note that the discrete times (\ref{Discrete tr}) at which the crack jumps are given by the recurrence formula
\begin{equation}\label{recurrence tr}
t_{j+1}=\dfrac{T_c}{q}\left[\left(a(t_j)+\ell\right)^p-A^p\right]\qquad {\rm with} \qquad a(t_0)=a(0)=A,
\end{equation}
$j=0,1,...,J-1,J$, where, again,  $a(t), p, q$ are given by relations (\ref{crack_evolution})$_1$. 

\begin{remark}\label{R_ell}
\emph{Physically, $\ell$ corresponds to the length over which the material is damaged around the crack fronts during the loading process, which is expected to correlate with the largest heterogeneity length scale of the material. For typical rubbers, this length scale is in the order of 1 micron \citep{Creton20}. For ceramics, on the other hand, it is typically in the order of 10 microns \citep{Ritchie99}. For other materials, this length scale could be even larger. For instance, for mortar, it is in the order of 1 millimeter \citep{Giaccio98}.}
\end{remark}

%
%

\section{A Griffith description of the growth of the crack in the elementary plate problem}\label{Sec: Griffith description}

Having established that for the elementary problem presented above the crack grows according to equation (\ref{crack_evolution}) in general, and having posited that, in particular, the crack grows discontinuously in time by jumps of length scale $\ell$, we now turn to working out how such a growth can be described by the type of Griffith formulation outlined in the Introduction.

\subsection{The Griffith criticality condition} 

Guided by experimental observations, we begin by postulating that the crack can only grow whenever the energy release rate $-\partial\mathcal{W}/\partial \Gamma$ reaches a critical value: 
\begin{equation*}
-\dfrac{\partial\mathcal{W}}{\partial \Gamma}=\mathcal{G}_c.
\end{equation*}

For the problem at hand, we recall that the energy release rate specializes (assuming, again, that the crack remains far away from the boundaries of the plate) to the classical result
\begin{equation}\label{ERR-Plate}
-\dfrac{\partial\mathcal{W}}{\partial \Gamma}=\dfrac{K^2_{\texttt{I}}}{E}=\dfrac{\left(\sigma(t)\sqrt{\pi a(t)}\right)^2}{E}.
\end{equation}

\subsection{The critical energy release rate $\mathcal{G}_c$ as a material function that evolves around the crack fronts} 

Next, also guided by experimental observations, we consider that  the critical energy release rate $\mathcal{G}_c$ of the material is \emph{not} a material constant, but rather a material function of both space and time that evolves from its initial uniform value 
\begin{equation*}
\mathcal{G}_c(\bfX,0)\equiv G_{c} \quad \textrm{for all}\quad \bfX\in\Omega
\end{equation*}
at time $t=0$ to  
\begin{equation*}
\mathcal{G}_c(\bfX,t)=\left(1-\theta_{\ell}(\bfX,t)\right)G_c+\theta_{\ell}(\bfX,t)D(\bfX,t)G_c,\quad (\bfX,t)\in\Omega\times (0,T],
\end{equation*}
as the loading progresses. Here, 
\begin{equation}\label{thetaD}
\theta_{\ell}(\bfX,t)=\left\{\begin{array}{ll}
1 & {\rm if}\quad \bfX\in\Omega_{\ell}(t) \vspace{0.2cm}\\
0 & {\rm else}\end{array}\right.
\end{equation}
stands for the characteristic or indicator function of the (collective) region $\Omega_{\ell}(t)$ where the value of the critical energy release rate evolves and, more specifically, decreases. We refer to $\Omega_{\ell}(t)$ as the degradation region. The evolution of the value of the critical energy release rate in this region is described by a degradation function $D(\bfX,t)$. For now, this function is required to satisfy the basic property
\begin{equation}\label{DX-properties}
0\leq D(\bfX,t)\leq 1.
\end{equation}

\subsubsection{The degradation region $\Omega_{\ell}(t)$} 

At this stage, to make further progress, we must specify the degradation region $\Omega_{\ell}(t)$ in (\ref{thetaD}). Initially, at time $t=0$, the location and the length scale of this region are clear. Indeed, the initial degradation region $\Omega_{\ell}(0)$ is comprised of two domains around the two crack fronts, both of which are of the length scale $\ell$ introduced above over which the material is initially damaged. As for the shape of these domains, at present, there is little experimental guidance. Our preliminary results indicate, however, that domains of different shapes lead essentially to the same results, so long as they have the same characteristic size $\ell$ and are equiaxed. Accordingly, we consider that the initial degradation region $\Omega_{\ell}(0)$ is comprised of two through-plate-thickness cylinders of diameter $\ell$ centered at $\pm (A+\ell/2)$ in the $\bfe_1$ axis, namely,
\begin{equation}\label{OmegaD-plate-0}
\Omega_{\ell}(0)=\left\{\bfX:\left(X_1\pm \left(A+\dfrac{\ell}{2}\right)\right)^2+X_2^2<\left(\dfrac{\ell}{2}\right)^2,\,|X_3|<\dfrac{B}{2}\right\}.
\end{equation}

As the crack fronts advance during the loading process, so does the degradation region $\Omega_{\ell}(t)$. In view of the inherent smallness of the material length scale $\ell$, we do not consider past domains where the degradation has taken place to be part of the degradation region $\Omega_{\ell}(t)$ at the current time $t\in(0,T]$. Instead, we consider that the degradation region $\Omega_{\ell}(t)$ is exclusively comprised of domains of length scale $\ell$ around the \emph{current} crack fronts. Precisely, much like for the initial degradation region (\ref{OmegaD-plate-0}), we  consider that the degradation region $\Omega_{\ell}(t)$ at the current time $t\in(0,T]$ is comprised of two through-plate-thickness cylinders of diameter $\ell$ centered at $(a(t)+\ell/2)$ in the $\bfe_1$ axis. That is,
\begin{equation}\label{OmegaD-plate}
\Omega_{\ell}(t)=\left\{\bfX:\left(X_1\pm \left(a(t)+\dfrac{\ell}{2}\right)\right)^2+X_2^2<\left(\dfrac{\ell}{2}\right)^2,\,|X_3|<\dfrac{B}{2}\right\}.
\end{equation}

\begin{remark}\label{Remark on No Intersection}
\emph{For the plate problem at hand, wherein the crack grows by jumps at time discrete values $t_j\in\{0=t_0,t_1,...,t_{J-1},t_{J}=T\}$, the degradation region (\ref{OmegaD-plate}) is such that
\begin{equation*}
\Omega_{\ell}(t)=\Omega_{\ell}(t_j),\qquad t_j\leq t<t_{j+1},\qquad {\rm with}\qquad \Omega_{\ell}(t_j)\bigcap\Omega_{\ell}(t_{j+1})=\emptyset
\end{equation*}
for all $j={0,1,...J-1,J}$.
}
\end{remark}

\subsubsection{The degradation function $D(\bfX,t)$} 

Having specified the degradation region (\ref{OmegaD-plate}), we can now proceed to specify additional properties for the degradation function $D(\bfX,t)$. Leveraging once more the inherent smallness of the material length scale $\ell$, it is natural to consider that the value of the critical energy release rate $\mathcal{G}_c(\bfX,t)$ is \emph{uniform in space} for $\bfX\in\Omega_{\ell}(t)$. In other words, it is natural to consider degradation functions $D(\bfX,t)$ that are independent of $\bfX$. With a slight abuse of notation, we write
\begin{equation}\label{D-X}
D(\bfX,t)=D(t).
\end{equation}
In view of Remark \ref{Remark on No Intersection} above, it is also natural to consider that there is no degradation in $\Omega_{\ell}(t)$ at $t=t_j$, right after the crack has jumped, and that the ensuing degradation in each $\Omega_{\ell}(t_j)$ should be monotonic. In short, when put together with the property (\ref{DX-properties}), we have 
\begin{equation}\label{Dt-properties}
0\leq D(t)\leq 1,\qquad D(t_j)=1,\qquad D(\tau_1)\geq D(\tau_2)\; \textrm{ for all } t_j<\tau_1<\tau_2< t_{j+1}
\end{equation}
for all time discrete values $t_j\in\{0=t_0,t_1,...,t_{J-1},t_{J}=T\}$ at which the crack jumps. 

The constitutive choices (\ref{D-X})-(\ref{Dt-properties}) --- together with the postulate that the crack grows only whenever the energy release rate (\ref{ERR-Plate}) reaches a critical value --- are entirely consistent with the premise that the crack grows via a discontinuous process in time by jumps of length scale $\ell$. Indeed, the constitutive choices (\ref{D-X})-(\ref{Dt-properties}) imply that from one instance in time $t_j$ to the next $t_{j+1}$, the crack will grow by jumping in an increment of size $\ell$, from $a(t_j)$ to $a(t_{j+1})=a(t_j)+\ell$, provided that
\begin{equation}\label{D-X-lim}
\displaystyle\lim_{t\nearrow t_{j+1}}D(t)G_c=\displaystyle\lim_{t\nearrow t_{j+1}}\dfrac{\left(\sigma(t)\sqrt{\pi a(t)}\right)^2}{E}=\dfrac{\left(\sigma(t_{j+1})\sqrt{\pi a(t_j)}\right)^2}{E}.
\end{equation}
That is, according to the Griffith criticality condition 
\begin{equation}\label{Gc-evol-Dt}
-\dfrac{\partial\mathcal{W}}{\partial \Gamma}=\dfrac{\left(\sigma(t)\sqrt{\pi a(t)}\right)^2}{E}=\left(1-\theta_{\ell}(\bfX,t)\right)G_c+\theta_{\ell}(\bfX,t)
D(t)G_c=\mathcal{G}_c(\bfX,t),
\end{equation}
any choice of degradation function $D(t)$ that satisfies the properties (\ref{Dt-properties}) and (\ref{D-X-lim}) describes \emph{exactly} any given crack growth (\ref{crack_evolution}) in the plate.

\begin{remark}\label{Remark on D}
\emph{Note that the condition (\ref{D-X-lim}) only requires the degradation function $D(t)$ to approach the value of $(\sigma(t_{j+1})\sqrt{\pi a(t_j)})^2/(G_c E)$ in the limit as $t\nearrow t_{j+1}$. It does \emph{not} place any restrictions on $D(t)$ otherwise.}
\end{remark}

%
\begin{figure}[t!]
  \subfigure[]{
   \begin{minipage}[]{0.48\linewidth}
   \centering \includegraphics[width=\linewidth]{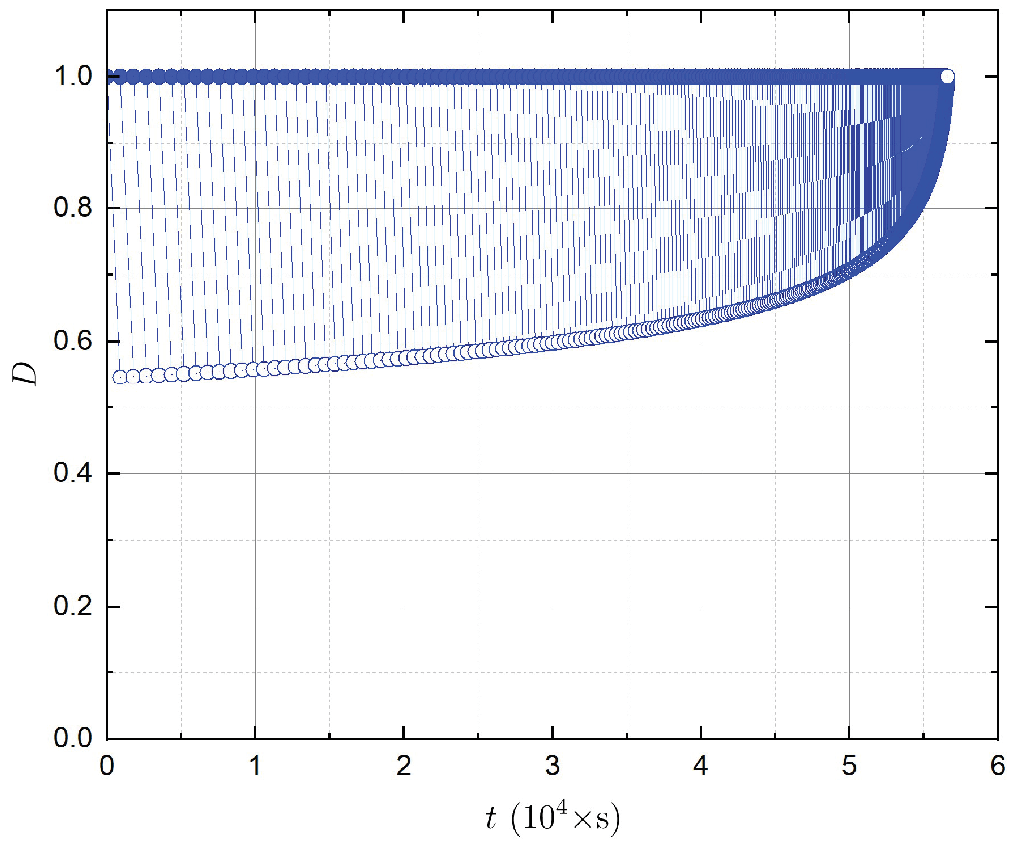}
   \end{minipage}}
  \subfigure[]{
   \begin{minipage}[]{0.48\linewidth}
   \centering \includegraphics[width=\linewidth]{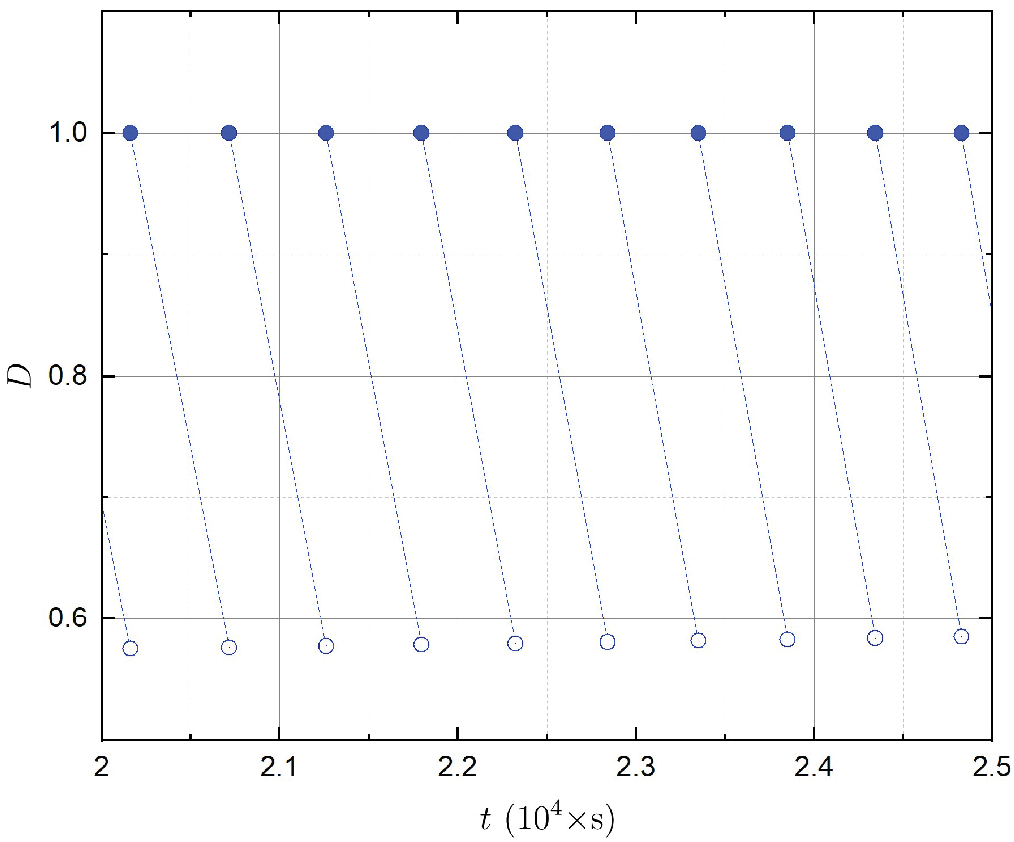}
   \end{minipage}}\par\centering
   \caption{The example (\ref{Dt-example}) of a piecewise linear degradation function $D(t)$  that leads to a Griffith criticality condition (\ref{Gc-evol-Dt}) that describes exactly the crack growth presented in Fig. \ref{Fig3} for a silicon nitride. (a) Plot of $D(t)$  over the entire loading interval $[0,5.6615\times10^4]$ s. (b)  Close-up over the interval $[2.00\times 10^4,2.50\times 10^4]$ s.}
   \label{Fig4}
\end{figure}
%
By way of an example, Fig. \ref{Fig4} provides the plot of a specific choice of degradation function $D(t)$ satisfying (\ref{Dt-properties})-(\ref{D-X-lim}) that describes the crack growth presented in Fig. \ref{Fig3} for a silicon nitride. In this case, $D(t)$ is chosen to be the piecewise linear function 
\begin{align}\label{Dt-example}
D(t)=\left\{\begin{array}{ll} 
1, & t_j=t \vspace{0.2cm}\\
1-\dfrac{1}{t_{j+1}-t_j}\left(1-\dfrac{\left(\sigma(t_{j+1})\sqrt{\pi a(t_j)}\right)^2}{G_c E}\right)(t-t_j), & t_j< t<t_{j+1}
\end{array}\right.,
\end{align}
where we recall that the applied stress $\sigma(t)$, the half size $a(t)$ of the crack, and the discrete times $t_j$ at which the crack jumps are given by the expressions (\ref{eq_triangluar_load}), (\ref{crack_evolution}), and (\ref{recurrence tr}), respectively. 

From Fig. \ref{Fig4} it is of note that the limiting values $\lim_{t\nearrow t_{j}}D(t)$ are ordered such that $\lim_{t\nearrow t_{j}}D(t)<\lim_{t\nearrow t_{j+1}}D(t)$. This is consistent with the expectation that, as the crack size increases, so does the energy release rate at the discrete times $t_j$ at which the crack jumps, and hence a smaller degradation of the critical energy release rate is sufficient for the crack to grow. 

\subsection{A constitutive prescription for $D(t)=\mathcal{D}(\mathfrak{g}(t))$ in terms of an internal memory variable $\mathfrak{g}(t)$}

As alluded to in Remark \ref{Remark on D} above, the properties (\ref{Dt-properties})-(\ref{D-X-lim}) are satisfied by a large class of functions. Given that the degradation function $D(t)$ describes the physical process by which the critical energy release rate $\mathcal{G}_c(\bfX,t)$ decreases around the crack fronts as a result of the large strains and large stresses that localize there, we posit that it can be described \emph{constitutively} in terms of some measure of the history of the elastic fields around the crack fronts. Precisely, we posit that there is a material degradation function $\mathcal{D}(\mathfrak{g})$ such that
\begin{align}\label{Dg-single}
D(t)=\mathcal{D}\left(\mathfrak{g}(t)\right),
\end{align}
where $\mathfrak{g}(t)$ stands for an appropriate measure of the history of the elastic fields, or internal memory variable, in the degradation region $\Omega_{\ell}(t)$. In this subsection, we introduce a particular form for this internal memory variable and the associated constitutive prescription (\ref{Dg-single}).

\subsubsection{The internal memory variable $\mathfrak{g}(t)$} 

We begin by considering internal memory variables within the large class
\begin{align}\label{g-tau}
\mathfrak{g}(t)=\left\{\begin{array}{ll} 
0, & t_j=t \vspace{0.2cm}\\
\displaystyle\int_{t_j}^t \mathcal{H}\left(f(\tau)\right)\mathcal{H}\left(g(\tau)\right)\mathcal{H}\left(\dot{g}(\tau)\right)\dot{g}(\tau)\,{\rm d}\tau, & t_j< t<t_{j+1}
\end{array}\right.,
\end{align}
where, again, $t_j\in\{0=t_0,t_1,...,t_{J-1},t_{J}=T\}$ stand for the time discrete values at which the crack jumps, and where $\mathcal{H}(\cdot)$ denotes the Heaviside function, the  ``dot'' notation stands for the material time derivative (i.e., with $\bfX$ fixed), and $f(t)$ and $g(t)$ are suitably selected functions of the elastic fields in the degradation region $\Omega_{\ell}(t)$. 

By definition, internal memory variables of the form (\ref{g-tau}) are non-negative functions of time, monotonically increasing in each time interval $t\in(t_j,t_{j+1})$,  that account for the loading history in a cumulative way. This accumulation --- which resets to $0$ at each $t_j\in\{0=t_0,t_1,...,t_{J-1},t_{J}=T\}$, right after the crack has jumped and a new degradation region $\Omega_{\ell}(t_j)$ has formed --- occurs only when $g(t)$ is both positive and increasing in time and, in addition, a condition of the type $f(t)>0$ is also satisfied.

\paragraph{The choice of functions $f(t)$ and $g(t)$ describing the elastic fields that lead to damage around the crack fronts} Based on experimental observations, the choice of functions $f(t)$ and $g(t)$ in (\ref{g-tau}) should be such that they:
\begin{itemize}

\item{are bounded, and}

\item{correlate with the local strains $\bfE(\bfu)$ --- or, equivalently, the local stresses $\boldsymbol{\sigma}(\bfX,t)$ --- in the degradation region $\Omega_{\ell}(t)$.}
    
\end{itemize}
The first of these requirements ensures that $f(t)$ and $g(t)$ can be viewed as signed measures. The second one ensures that these measures increase when the local strains --- or, equivalently, the local stresses --- increase, consistent with experimental observations.

In view of the above two requirements, arguably the simplest and, at the same time, the most natural choices for $f(t)$ and $g(t)$ are the average of the trace of the strain ${\rm tr}\,\bfE(\bfu)$ over the degradation region $\Omega_{\ell}(t)$ and the average of the stored-energy function $W(\bfE(\bfu))$ over the degradation region $\Omega_{\ell}(t)$, normalized by the new area of the crack that would be created in that region, say $\mathcal{A}_{\ell}$, namely, 
\begin{align}\label{f(t)-def}
f(t)=\displaystyle\int_{\Omega_{\ell}(t)}{\rm tr}\,\bfE(\bfu)\,{\rm d}\bfX
\end{align}
and
\begin{align}\label{g(t)-def}
g(t)=\dfrac{1}{\mathcal{A}_{\ell}}\displaystyle\int_{\Omega_{\ell}(t)}W(\bfE(\bfu))\,{\rm d}\bfX.
\end{align}

\begin{remark}
\emph{The choice (\ref{f(t)-def}) for the function $f(t)$ in the internal memory variable (\ref{g-tau}) allows to distinguish between loadings for which the volume around crack fronts increases ($f(t)>0$) from those for which the volume decreases ($f(t)<0$). Damage around crack fronts is known to occur mostly, if not exclusively, when the volume of these regions increases.}
\end{remark}

\begin{remark}
\emph{The choice (\ref{g(t)-def}) for the function $g(t)$ in the internal memory variable (\ref{g-tau}) provides a measure of the strains that localize around crack fronts. Damage around crack fronts is known to increase when the strains --- and hence the elastic energy --- in these regions increase.}
\end{remark}

\begin{remark}
\emph{The normalization by the added crack area $\mathcal{A}_{\ell}$ in (\ref{g(t)-def}) yields a quantity of units $force/length$, exactly as those of the critical energy release rate $\mathcal{G}_c(\bfX,t)$.}
\end{remark}

\begin{remark}
\emph{The internal memory variable (\ref{g-tau}) with (\ref{f(t)-def})-(\ref{g(t)-def}) is fundamentally different from the types of internal memory variables that have been recently used in the phase-field models of fatigue mentioned in the Introduction in that it is \emph{not} a pointwise variable taking different values at different material points $\bfX$, instead, it is a volume average over the degradation region $\Omega_\ell(t)$.}
\end{remark}

\paragraph{The specialization of the internal memory variable (\ref{g-tau}) with (\ref{f(t)-def})-(\ref{g(t)-def}) to the plate problem at hand} For the elementary plate problem of interest here, given that the displacement field $\bfu(\bfX,t)$ --- and hence the strain field $\bfE(\bfu)$ and the stress field $\boldsymbol{\sigma}(\bfX,t)$ --- is known explicitly from the classical work of \cite{Westergaard39}, and given that the degradation region $\Omega_{\ell}(t)$ is described by (\ref{OmegaD-plate}), which implies that $\mathcal{A}_{\ell}=\ell B$, we can readily determine the associated trace of the strain
\begin{align*}
{\rm tr}\,\bfE(\bfu)=\dfrac{(1-2\nu)}{E}\left(\dfrac{2}{a(t)}{\rm Re}\left[\frac{Z}{\sqrt{\frac{Z^2}{a^2(t)}-1}}\right] -1\right)\sigma(t)
\end{align*}
and the stored-energy function 
\begin{align*}
W(\bfE(\bfu))=&\dfrac{\sigma^2(t)}{2E}\left(1+\frac{2 (1-\nu)}{a^2(t)}
   {\rm Re}\left[\frac{Z}{\sqrt{\frac{Z^2}{a^2(t)}-1}}\right] {\rm Re}\left[\frac{Z}{\sqrt{\frac{Z^2}{a^2(t)}-1}}-a(t)\right]-\right.\\
   &\frac{2 (1+\nu)}{a(t)} {\rm Im}\left[\left(\frac{Z^2}{a^2(t)}-1\right)^{-3/2}\right]X_2+\\
   &\left.\frac{2 (1+\nu)}{a^2(t)}
   \left({\rm Im}\left[\left(\frac{Z^2}{a^2(t)}-1\right)^{-3/2}\right]^2+{\rm Re}\left[\left(\frac{Z^2}{a^2(t)}-1\right)^{-3/2}\right]^2\right)X_2^2\right),
\end{align*}
where $Z=X_1+i X_2$, and, in turn, we can readily carry out the integrals in (\ref{f(t)-def}) and (\ref{g(t)-def}). The results read
\begin{align}\label{fr-Plate}
f(t)=\dfrac{(1-2\nu)\pi \ell^{3/2}}{16 E a^{3/2}(t)}\left(8-\dfrac{4\sqrt{\ell}}{\sqrt{a(t)}}+\dfrac{3\ell}{a(t)}\right)B a^2(t)\sigma(t)
\end{align}
and
\begin{align}\label{gr-Plate}
g(t)=k(t)\dfrac{\left(\sigma(t)\sqrt{\pi a(t)}\right)^2}{E}\quad {\rm with}\quad k(t)=\dfrac{1}{6\pi}+\left(\dfrac{5}{12\pi}+\dfrac{1}{8}\right)(1-\nu)+O\left(\left(\dfrac{\ell}{a(t)}\right)^{1/2}\right).
\end{align}

\begin{remark}
\emph{The result (\ref{fr-Plate}) implies that $\mathcal{H}(f(t))=\mathcal{H}(\sigma(t))$, independent of $\Omega_\ell(t)$.}
\end{remark}

\begin{remark}
\emph{The result (\ref{gr-Plate}) happens to be nothing more than a multiple of the energy release rate (\ref{ERR-Plate}). This connection will become more apparent in the next section.}
\end{remark}

Upon substituting the expressions (\ref{fr-Plate}) and (\ref{gr-Plate}) in (\ref{g-tau}) and carrying out the resulting integral, we obtain the explicit expression 
\begin{align}\label{g-tau-Plate}
\mathfrak{g}(t)=\left\{\begin{array}{ll} 
0, & t_j=t \vspace{0.2cm}\\
\left(\left\lfloor\dfrac{t}{T_c}\right\rfloor-\left\lfloor\dfrac{t_j}{T_c}\right\rfloor\right)g_{max}(t_j)  + c_l\left(t\right)- c_l\left(t_j\right), & t_j< t<t_{j+1} \end{array}\right. ,
\end{align}
where $\lfloor \cdot\rfloor$ stands for the floor function,
\begin{align*}
c_l\left(t\right)=\left\{\begin{array}{ll} 
4\left(\left\lfloor\dfrac{t}{T_c}\right\rfloor-\dfrac{t}{T_c}\right)^2g_{max}(t_j), & \left\lfloor\dfrac{t}{T_c}\right\rfloor T_c\leq t\leq \left\lfloor\dfrac{t}{T_c}\right\rfloor T_c + \dfrac{T_c}{2} \vspace{0.2cm}\\
g_{max}(t_j), & \left\lfloor\dfrac{t}{T_c}\right\rfloor T_c + \dfrac{T_c}{2}\leq t\leq \left(\left\lfloor\dfrac{t}{T_c}\right\rfloor +1\right) T_c  \end{array}\right. ,
\end{align*}
and
\begin{align*}
g_{max}(t_j)=k(t_j)\dfrac{\left(\sigma_{max}\sqrt{\pi a(t_j)}\right)^2}{E}\quad {\rm with}\quad k(t_j)=\dfrac{1}{6\pi}+\left(\dfrac{5}{12\pi}+\dfrac{1}{8}\right)(1-\nu)+O\left(\left(\dfrac{\ell}{a(t_j)}\right)^{1/2}\right)
\end{align*}
for the internal memory variable.

%
\begin{figure}[t!]
  \subfigure[]{
   \begin{minipage}[]{0.48\linewidth}
   \centering \includegraphics[width=\linewidth]{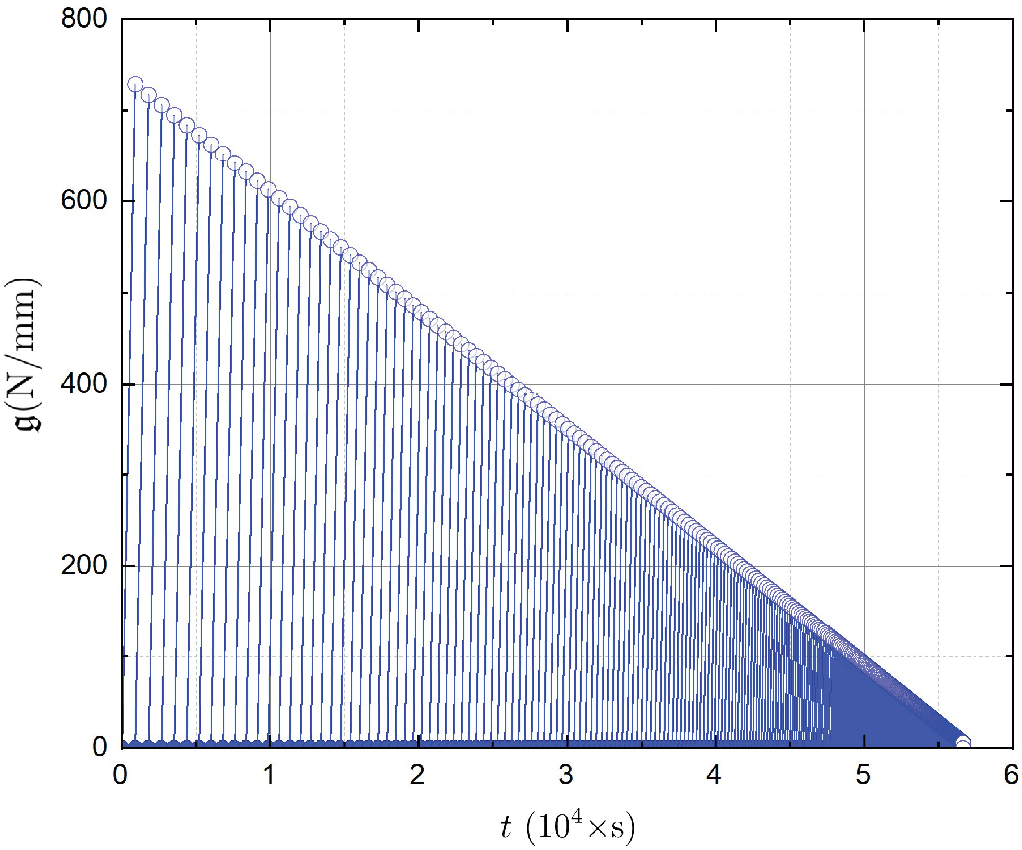}
   \end{minipage}}
  \subfigure[]{
   \begin{minipage}[]{0.48\linewidth}
   \centering \includegraphics[width=\linewidth]{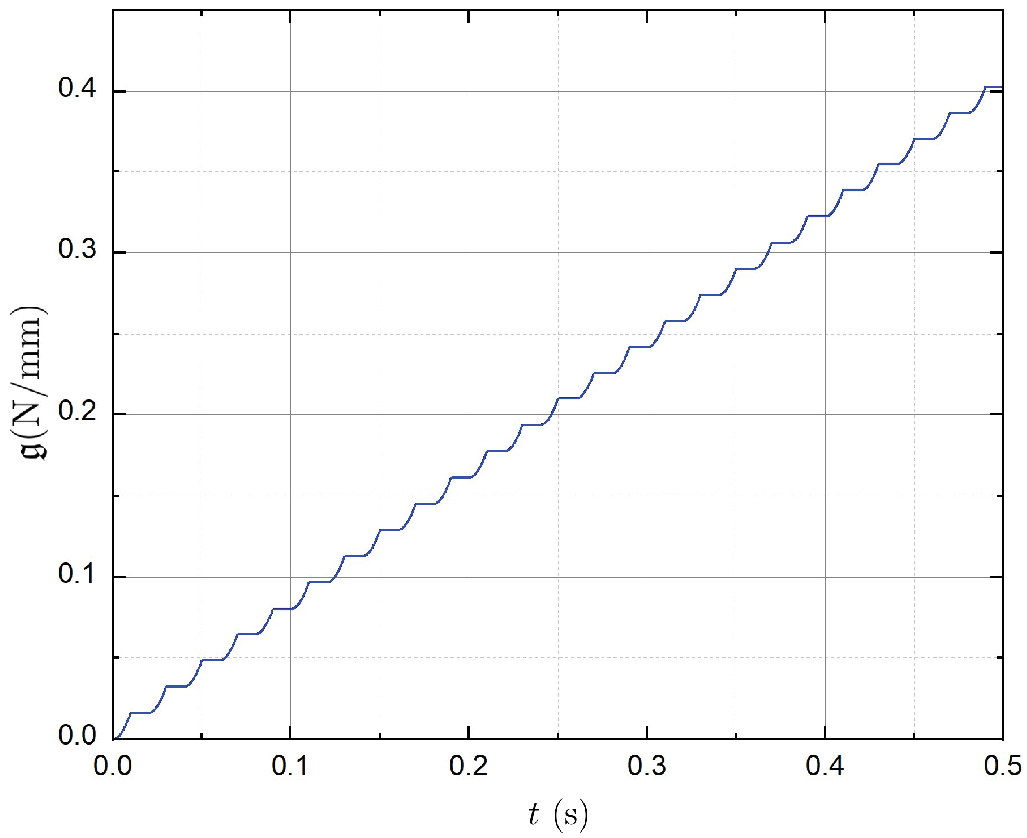}
   \end{minipage}}\par\centering
   \caption{Evolution of the internal memory variable (\ref{g-tau-Plate}) for the crack growth in a silicon nitride presented in Fig. \ref{Fig3}. (a) Plot of $\mathfrak{g}(t)$  over the entire loading interval $[0,5.6615\times10^4]$ s. (b)  Close-up over the interval $[0,0.5]$ s.}
   \label{Fig5}
\end{figure}
%
By way of an example, Fig. \ref{Fig5}  provides a plot of the internal memory variable (\ref{g-tau-Plate}) for the crack growth presented in Fig. \ref{Fig3}. That is, the result pertains to the case of a crack of initial half size $A=10$ mm, loading parameters, $\sigma_{max}=25$ MPa, $T_c=0.02$ s, $T=T_{\texttt{m}}=5.6615\times 10^4$ s, and material parameters $E=300$ GPa, $\nu=0.25$, $G_c=120$ N/m, $\ell=20$ $\mu$m. 

As dictated by the definition (\ref{g-tau}), one key characteristic that is immediate from  Fig. \ref{Fig5} is that the internal memory variable $\mathfrak{g}(t)$ is indeed a non-negative function of time that increases monotonically in each time interval $t\in(t_j,t_{j+1})$. Another immediate key characteristic is that $\mathfrak{g}(t)$ is markedly nonlinear, as a result of its growth only taking place when $g(t)> 0$ and $\dot{g}(t)> 0$.

\subsubsection{The material degradation function $\mathcal{D}\left(\mathfrak{g}\right)$}\label{Sec: Dg-cali}

Having defined the internal memory variable (\ref{g-tau}) with (\ref{f(t)-def})-(\ref{g(t)-def}), and having worked out its specialization (\ref{g-tau-Plate}) to the plate problem of interest here, we are now in a position to define the material degradation function $\mathcal{D}\left(\mathfrak{g}\right)$ in (\ref{Dg-single}). 

Indeed, note that from the properties (\ref{Dt-properties})$_{1,2}$ and (\ref{D-X-lim}) of $D(t)$, we have the requirements
\begin{equation}\label{Dg-features}
0\leq \mathcal{D}\left(\mathfrak{g}\right)\leq 1,\quad \mathcal{D}\left(0\right)=1,\quad \displaystyle\lim_{t\nearrow t_{j+1}}\mathcal{D}\left(\mathfrak{g}(t)\right)=\dfrac{\left(\sigma(t_{j+1})\sqrt{\pi a(t_j)}\right)^2}{G_c E}
\end{equation}
on $\mathcal{D}\left(\mathfrak{g}\right)$. The third of these requirements is one that directly provides the values of the material degradation function $\mathcal{D}\left(\mathfrak{g}\right)$ at distinct values of its argument $\mathfrak{g}$. 

%
\begin{figure}[t!]
  \subfigure[]{
   \begin{minipage}[]{0.48\linewidth}
   \centering \includegraphics[width=\linewidth]{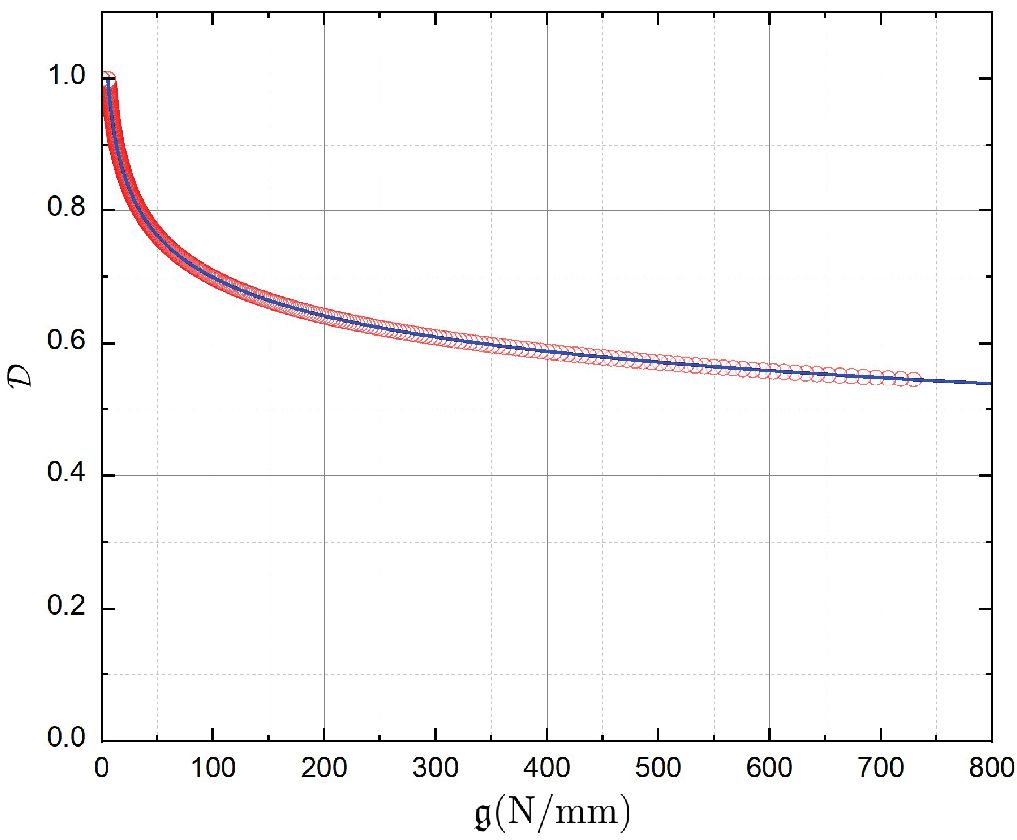}
   \end{minipage}}
  \subfigure[]{
   \begin{minipage}[]{0.48\linewidth}
   \centering \includegraphics[width=\linewidth]{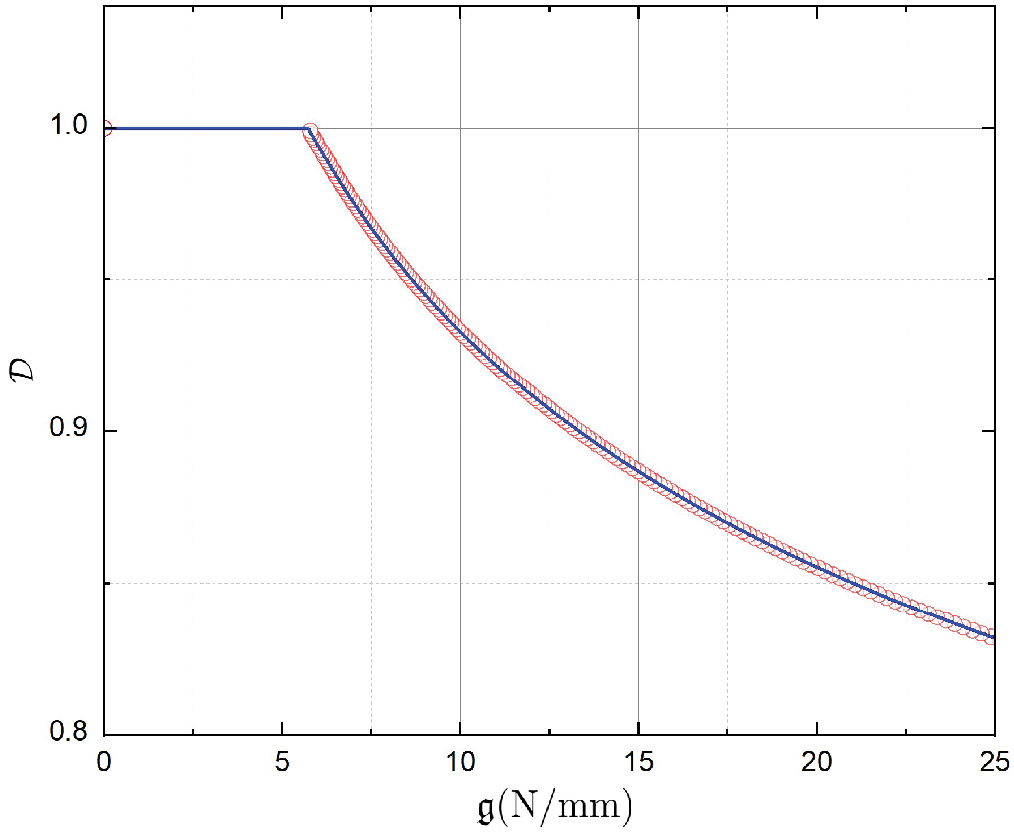}
   \end{minipage}}\par\centering
   \caption{Values (circles) of the degradation function $\mathcal{D}(\mathfrak{g})$ --- obtained from equations (\ref{Dg-features})$_2$ and (\ref{Dg-features})$_3$ --- in the Griffith criticality condition (\ref{Gc-evol-Dg-Plate}) that describe exactly the crack growth presented in Fig. \ref{Fig3}. For direct comparison, the formula (\ref{D-constitutive-formula}) with $\mathfrak{g}_{th}=5.734$ N/mm, $G_{\infty}/G_c=0$, $K=326$ (m/N)$^{\alpha}$, $\alpha=0.2112$, and $\beta=0.0861$ is also plotted (solid line). (a) Plot of $\mathcal{D}(\mathfrak{g})$  over the entire range of $\mathfrak{g}$ values $[0,730\times10^3]$ N/m. (b)  Close-up around $\mathfrak{g}=\mathfrak{g}_{th}$.}
   \label{Fig6}
\end{figure}
%

By way of an example, Fig. \ref{Fig6}  plots the values (circles) of the material degradation function $\mathcal{D}\left(\mathfrak{g}\right)$ generated by equations (\ref{Dg-features})$_2$ and (\ref{Dg-features})$_3$ for the case of the crack growth presented in Fig. \ref{Fig3} above. Two key observations from this figure are that the function $\mathcal{D}\left(\mathfrak{g}\right)$ is monotonically decreasing with respect to its argument $\mathfrak{g}$ and that there is a threshold value, say $\mathfrak{g}_{th}$, below which $\mathcal{D}\left(\mathfrak{g}\right)=1$. The presence of such a threshold is consistent with the experimental observation that there are loading regimes that do not cause enough damage to lead to a decrease in the value of the critical energy release rate.

For later reference, another observation worth remarking from the representative result shown in Fig. \ref{Fig6} is that the function  $\mathcal{D}\left(\mathfrak{g}\right)$ may be described by a formula of the form

\begin{align}\label{D-constitutive-formula}
\mathcal{D}(\mathfrak{g})=\left\{\begin{array}{ll} 
1, & \mathfrak{g}\leq \mathfrak{g}_{th}\vspace{0.2cm}\\
\dfrac{G_{\infty}}{G_c}+\dfrac{1-\dfrac{G_{\infty}}{G_c}+\left(1-\dfrac{G_\infty}{G_c}\right)K\mathfrak{g}_{th}^{\alpha} \left(\dfrac{\mathfrak{g}}{\mathfrak{g}_{th}}\right)^{\beta}}{1+ K \mathfrak{g}^{\alpha}}, & \mathfrak{g}>\mathfrak{g}_{th} \end{array}\right. ,
\end{align}
where $\mathfrak{g}_{th}\geq0$, $G_{\infty}\geq0$, $K\geq0$, and $\alpha>\beta\geq0$ are material constants. For direct comparison, Fig. \ref{Fig6} includes the plot (solid line) of this formula for the constants $\mathfrak{g}_{th}=5.734$ N/mm, $G_{\infty}/G_c=0$, $K=326$ (m/N)$^{\alpha}$, $\alpha=0.2112$, $\beta=0.0861$. Incidentally, the formula (\ref{D-constitutive-formula}) is of a form that has been repeatedly used in the literature to model a wide variety of dissipative phenomena ranging from the polarization of dielectrics \citep{Cole41} to the deformation of viscous fluids and viscoelastic solids \citep{Cross65,KLP16}.

Summing up, the above derivation has shown that the Griffith criticality condition 
\begin{equation}\label{Gc-evol-Dg-Plate}
-\dfrac{\partial\mathcal{W}}{\partial \Gamma}=\dfrac{\left(\sigma(t)\sqrt{\pi a(t)}\right)^2}{E}=\left(1-\theta_{\ell}(\bfX,t)\right)G_c+\theta_{\ell}(\bfX,t)
\mathcal{D}\left(\mathfrak{g}(t)\right)G_c=\mathcal{G}_c(\bfX,t),
\end{equation}
where $\theta_{\ell}(\bfX,t)$, given by (\ref{thetaD}), is the indicator function of the degradation region (\ref{OmegaD-plate}), the internal memory variable $\mathfrak{g}(t)$ is given by (\ref{g-tau-Plate}), and $\mathcal{D}\left(\mathfrak{g}\right)$ is any material function that satisfies (\ref{Dg-features}), e.g., a function of the form (\ref{D-constitutive-formula}), describes \emph{exactly} any given crack growth (\ref{crack_evolution}) in the elementary plate problem presented in Section \ref{Sec: Plate}, this for any initial crack half size $A$, constants $C$, $m$, and crack jump characteristic size $\ell$, any Young's modulus $E$, Poisson's ratio $\nu$, and initial critical energy release rate $G_c$, and any maximum stress $\sigma_{max}$ and cycle duration $T_c$.

\section{The proposed Griffith formulation of crack growth for non-monotonic loading}\label{Sec: General}

The ideas introduced in the preceding section that culminated in the Griffith criticality condition (\ref{Gc-evol-Dg-Plate}) to describe the growth of the crack in the elementary plate problem presented in Section \ref{Sec: Plate} can be seamlessly generalized to describe the growth of large cracks in bodies of arbitrary geometry (not just a plate containing a centered crack), made of a linear elastic brittle material, that are subjected to arbitrary non-monotonic quasistatic loading conditions (not just cyclic tensile loading). As we elaborate in this section, the generalization hinges on the observation that the material degradation function $\mathcal{D}\left(\mathfrak{g}\right)$ that describes the decrease of the critical energy release rate around crack fronts is an \emph{intrinsic macroscopic material property}, one that is entirely determined by the history of the strain field around crack fronts, whatever the geometry of the body and whatever the applied loading conditions may be. 

\begin{figure}[t!]
\centering
\centering\includegraphics[width=0.85\linewidth]{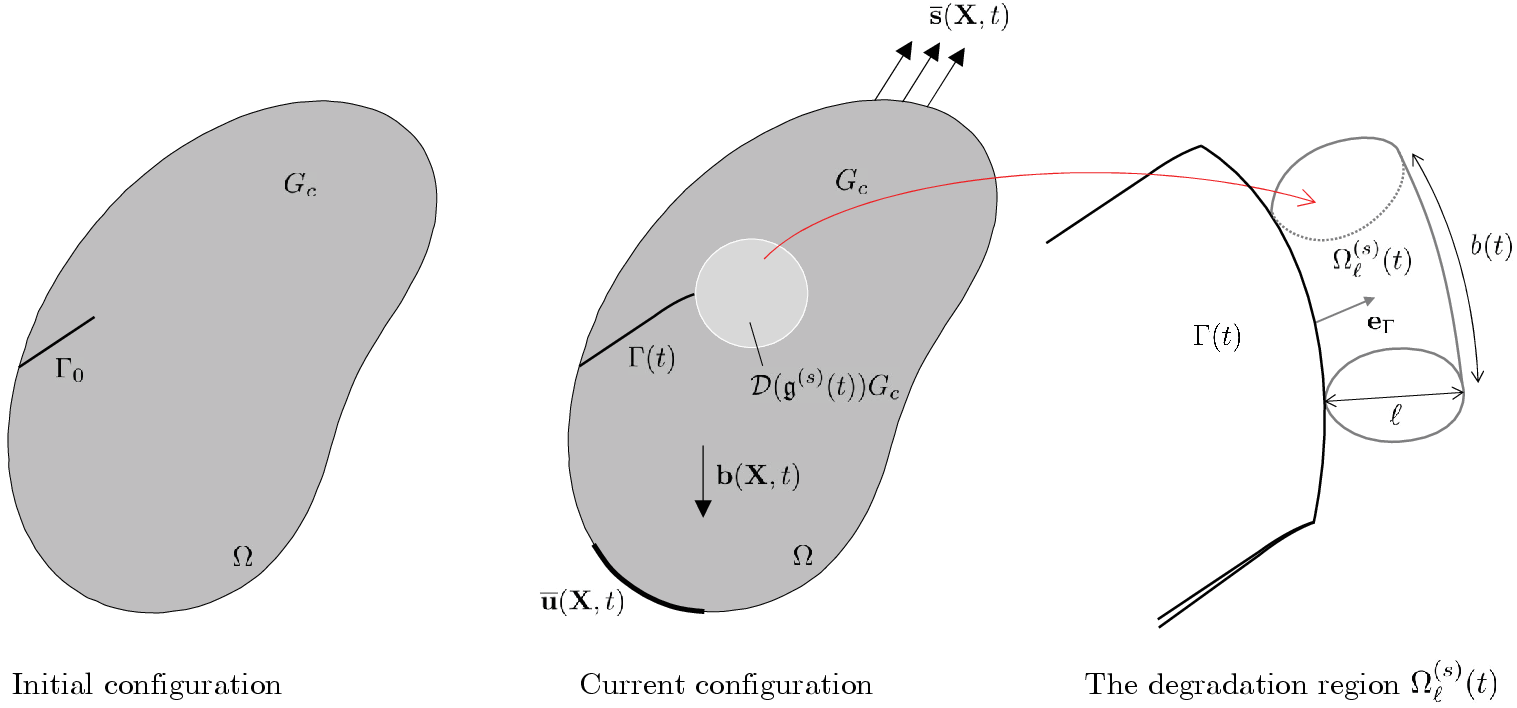}
\caption{\small Schematic of the growth of a large crack in a body made of a nominally elastic brittle material that is subjected to a surface force $\overline{\bfs}(\bfX,t)$ and displacement $\overline{\bfu}(\bfX,t)$ on its boundary and a body force $\textbf{b}(\bfX,t)$ in its volume, all of which are arbitrary non-monotonic functions of time $t$. The schematic also illustrates one of the underlying degradation regions $\Omega^{(s)}_{\ell}(t)$ --- a tube of constant diameter $\ell$ and arc length $\mathit{b}(t)$ that emanates from the crack front, in the direction $\bfe_{\Gamma}$ of the crack front --- wherein the critical energy release rate $\mathcal{G}_c$ decreases from its initial value $G_c$ to $\mathcal{D}(\mathfrak{g}^{(s)}(t))G_c\leq G_c$, as dictated by the material degradation function $\mathcal{D}(\mathfrak{g})$ in terms of the internal memory variable $\mathfrak{g}^{(s)}(t)$.}\label{Fig7}
\end{figure}

Consider thus a body that initially, at time $t=0$, occupies a bounded open domain $\Omega\subset \mathbb{R}^3$ that contains a large pre-existing crack\footnote{For clarity of exposition, we restrict attention to the presence of a single crack in the body. The formulation for multiple cracks follows readily from that of a single crack.} $\Gamma_0\subset\Omega$. The body is made of a linear elastic brittle material with stored-energy function $W(\bfE(\bfu))$. A part of its boundary $\partial\Omega_{\texttt{d}}$ is subjected to a prescribed displacement $\overline{\bfu}(\bfX,t)$, while the complementary part $\partial\Omega_{\texttt{s}}=\partial\Omega\setminus\partial\Omega_{\texttt{d}}$ is subjected to a prescribed force $\overline{\bfs}(\bfX,t)$ per unit undeformed area. The body is also subjected to a force $\textbf{b}(\bfX,t)$ per unit undeformed volume; see Fig. \ref{Fig7}. Importantly, the dependence on time $t\in[0,T]$ of the prescribed boundary data $\overline{\bfu}(\bfX,t)$ and $\overline{\bfs}(\bfX,t)$ and of the prescribed body force $\textbf{b}(\bfX,t)$ is arbitrary, with the proviso that they are applied slowly enough that inertial effects can be neglected. 

In response to the applied boundary conditions and body force, the body will deform and the crack may also possibly grow. We denote the crack at the current time $t$ by $\Gamma(t)\subset \Omega$. In this work, much like in the original formulation of \cite{Griffith21}, attention is restricted to the special case when the crack path is known. The more general case when the crack path is not known, but rather is determined from the formulation itself, amounts to generalizing the variational theory of \cite{Francfort98} to account for a critical energy release rate that is a material function and not just a material constant. We shall deal with such a generalization in future work.

Now, generalizing, \emph{mutatis mutandis}, the Griffith criticality condition (\ref{Gc-evol-Dg-Plate}) derived for the elementary plate problem above, we posit that the crack grows whenever the Griffith criticality condition 
\begin{equation}\label{Griffith-General}
-\dfrac{\partial\mathcal{W}}{\partial \Gamma}=\mathcal{G}_c(\bfX,t)=\left(1-\sum_{s=1}^S\theta^{(s)}_{\ell}(\bfX,t)\right)G_c+\sum_{s=1}^S\theta^{(s)}_{\ell}(\bfX,t)\mathcal{D}\left(\mathfrak{g}^{(s)}(t)\right)G_c
\end{equation}
is satisfied. Here,
\begin{equation*}
\mathcal{W}=\displaystyle\int_{\Omega\setminus\Gamma(t)}W(\bfE(\bfu))\,{\rm d}\bfX-\displaystyle\int_{\Omega\setminus\Gamma(t)}\textbf{b}\cdot\bfu\,{\rm d}\bfX-\displaystyle\int_{\partial\Omega_{\texttt{s}}}\overline{\bfs}\cdot\bfu\,{\rm d}\bfX
\end{equation*}
stands for the potential energy evaluated at the displacement field $\bfu(\bfX,t)$ that satisfies the equations of elastostatics, the derivative $-\partial\mathcal{W}/\partial \Gamma$ is to be carried out under fixed boundary conditions and fixed body force, 
$$G_c$$
is the initial uniform value of the critical energy release rate of the material, while 
\begin{equation*}
\left\{\begin{array}{l}
\theta^{(s)}_{\ell}(\bfX,t)\vspace{0.2cm}\\
\mathfrak{g}^{(s)}(t)\vspace{0.2cm}\\
\mathcal{D}\left(\mathfrak{g}\right)\end{array}\right.
\end{equation*}
are the functions that describe where, when, and how much the critical energy release rate $\mathcal{G}_c(\bfX,t)$ evolves and, in particular, decreases from $G_c$ to $\mathcal{D}\left(\mathfrak{g}^{(s)}(t)\right)G_c$ within the body. We now spell out each of these three types of functions, one at a time.

\paragraph{The degradation regions $\Omega^{(s)}_\ell(t)$ and their indicator functions $\theta^{(s)}_{\ell}(\bfX,t)$} Precisely,  
\begin{equation*}
\theta^{(s)}_{\ell}(\bfX,t)=\left\{\begin{array}{ll}
1 & {\rm if}\quad \,\bfX\in\Omega^{(s)}_{\ell}(t) \vspace{0.2cm}\\
0 & {\rm else}\end{array}\right.\qquad s=1,2,...,S-1,S
\end{equation*}
stand for the indicator functions of each degradation region $\Omega^{(s)}_{\ell}(t)$ wherein the value of the critical energy release rate decreases. As schematically depicted in Fig. \ref{Fig7}, consistent with the degradation region (\ref{OmegaD-plate}) introduced in the elementary plate problem, the degradation regions $\Omega^{(s)}_{\ell}(t)$ are identified as tubes of the same constant diameter $\ell$ and the same arc length $\mathit{b}(t)=B(t)/\lfloor B(t)/\ell\rfloor$ that emanate from the \emph{current} front of the crack $\Gamma(t)$, in the direction $\bfe_{\Gamma}$ of the crack front, and that their union $\Omega_\ell(t)=\bigcup_{s=1}^S\Omega^{(s)}_\ell(t)$ spans the entire arc length $B(t)$ of the crack front. We refer to $\Omega_\ell(t)$ as the collective degradation region.

\begin{remark}
\emph{Given a crack front of arc length $B(t)$, there is a total of $S=\lfloor B(t)/\ell\rfloor$ degradation regions $\Omega^{(s)}_{\ell}(t)$. In view of the bounds
\begin{equation*}
\ell\leq b(t)=\dfrac{B(t)}{\left\lfloor\dfrac{B(t)}{\ell}\right\rfloor}\leq 2\ell
\end{equation*}
on their arc length $b(t)$ and in view of the smallness of the characteristic size $\ell$, the spatial arrangement of the degradation regions $\Omega^{(s)}_{\ell}(t)$, $s=1,2,...,S-1,S$, along the crack front can be chosen arbitrarily. 
}
\end{remark}

\begin{remark}
\emph{For problems in which all material points along the crack front are subjected to the same strain --- as is the case, for instance, in any 2D problem --- it suffices to set $\mathit{b}(t)=B(t)$, that is, it suffices to define a sole degradation region $\Omega_\ell(t)$ of constant diameter $\ell$ and arc length equal to the entire arc length $B(t)$ of the crack front. Accordingly, in all the 2D plane-stress problems that we consider in the sequel, we set $\mathit{b}(t)=B(t)$ and drop the use of the superscript $(s)$ in $\theta_{\ell}(\bfX,t)$ and $\mathfrak{g}(t)$, so that the Griffith criticality condition (\ref{Griffith-General}) can be rewritten more compactly as
\begin{equation}\label{Griffith-General-2D}
-\dfrac{\partial\mathcal{W}}{\partial \Gamma}=\mathcal{G}_c(\bfX,t)=\left(1-\theta_{\ell}(\bfX,t)\right)G_c+\theta_{\ell}(\bfX,t)\mathcal{D}\left(\mathfrak{g}(t)\right)G_c.
\end{equation}
}
\end{remark}

\begin{remark}
\emph{As elaborated in Subsection \ref{Sec: ell sensitivity} below, the Griffith criticality condition (\ref{Griffith-General}) is largely insensitive to the precise value of the material length scale $\ell$, so long as it is selected not to exceed the correct order of magnitude, that is, the same order of magnitude as the length scale over which the material is damaged around crack fronts. Recalling from Remark \ref{R_ell} above that $\ell$ correlates with the size of the largest heterogeneity of the material at hand, it suffices to choose the value of $\ell$ based on such a size.}
\end{remark}

\paragraph{The internal memory variable $\mathfrak{g}^{(s)}(t)$} Consistent with the internal memory variable (\ref{g-tau}) with (\ref{f(t)-def})-(\ref{g(t)-def}) introduced in the elementary plate problem,
\begin{align}\label{g-tau-Gen}
\mathfrak{g}^{(s)}(t)=\left\{\begin{array}{ll} 
0, & t_j=t \vspace{0.2cm}\\
\displaystyle\int_{t_j}^t \mathcal{H}\left(\overline{I}^{(s)}_1(\tau)\right)\mathcal{H}\left(\dot{\overline{W}}^{(s)}(\tau)\right)\dot{\overline{W}}^{(s)}(\tau)\,{\rm d}\tau, & t_j< t<t_{j+1}
\end{array}\right.
\end{align}
stands for the internal memory variable that keeps track of the history of the strain field in the degradation region $\Omega^{(s)}_{\ell}(t)$. In this last expression, $t_j\in[0,T]$ denotes each instance in time at which the crack front advances during the prescribed loading, 
\begin{align}\label{I1-W-def}
\overline{I}^{(s)}_1(t):=\displaystyle\int_{\Omega^{(s)}_{\ell}(t)}{\rm tr}\,\bfE(\bfu)\,{\rm d}\bfX\qquad {\rm and}\qquad \overline{W}^{(s)}(t):=\dfrac{1}{\ell b(t)}\displaystyle\int_{\Omega^{(s)}_{\ell}(t)}W(\bfE(\bfu))\,{\rm d}\bfX.
\end{align}

\begin{remark}
\emph{By definition, the value of the internal memory variable (\ref{g-tau-Gen}) resets to zero every time that the crack front in $\Omega^{(s)}_\ell(t)$ advances at $t=t_j$. A corollary of this property is that the internal memory variable (\ref{g-tau-Gen}) remains identically zero for loading conditions for which the crack grows continuously in time.}
\end{remark}

\begin{remark}
\emph{For initial-boundary-value problems for which the stress field $\boldsymbol{\sigma}(\bfX,t)$ --- and hence the strain field $\bfE(\bfu)=(1+\nu)/E\,\boldsymbol{\sigma}-\nu/E({\rm tr}\boldsymbol{\sigma})\bfI$ --- happens to be known asymptotically around the crack front, the volume average quantities (\ref{I1-W-def}) can be determined analytically in terms of that asymptotic field on account of the smallness of $\ell$. In other words, analytical knowledge of the stress intensity factors for a given initial-boundary-value problem implies analytical knowledge (approximately, in an asymptotic sense) of the volume average quantities (\ref{I1-W-def}) for that initial-boundary-value problem. For example, for plane-stress problems for which a crack in a plate of constant thickness $B$ is subject to Mode $\texttt{I}$, use in (\ref{I1-W-def}) of the classical asymptotic stress field \citep{Irwin1958}
\begin{align*}
\boldsymbol{\sigma}(\bfX,t)=\dfrac{K_{\texttt{I}}(t)}{\sqrt{2\pi r}}\bfS(\theta)=\dfrac{K_{\texttt{I}}(t)}{\sqrt{2\pi r}}\cos\dfrac{\theta}{2}&\left[\left(1-\sin\dfrac{\theta}{2}\sin\dfrac{3\theta}{2}\right)\bfe_1\otimes\bfe_1+
\left(1+\sin\dfrac{\theta}{2}\sin\dfrac{3\theta}{2}\right)\bfe_2\otimes\bfe_2+\right.\\
&\left.\sin\dfrac{\theta}{2}\cos\dfrac{3\theta}{2}\left(\bfe_1\otimes\bfe_2+\bfe_2\otimes\bfe_1\right)\right]
\end{align*}
leads to the volume average quantities
\begin{equation}\label{ModeI-I1-W}
\overline{I}_1(t)=\left(\dfrac{1}{2}-\nu\right)\sqrt{\pi}B\ell^{3/2}\dfrac{K_{\texttt{I}}(t)}{E}\qquad {\rm and}\qquad \overline{W}(t)=k_0\dfrac{K^2_{\texttt{I}}(t)}{E}\quad {\rm with}\quad k_0=\dfrac{1}{6\pi}+\left(\dfrac{5}{12\pi}+\dfrac{1}{8}\right)(1-\nu).
\end{equation}
In these expressions, $r$ is the distance to the crack front, $\theta$ is the associated polar angle, and we have included the argument of time $t$ in the stress intensity factor $K_{\texttt{I}}(t)$ to make explicit its dependence on the evolving applied loads. The results (\ref{ModeI-I1-W}) make it plane that if $K_{\texttt{I}}(t)$ is known analytically, so are the volume average quantities $\overline{I}_1(t)$ and $\overline{W}(t)$.
}
\end{remark}

\paragraph{The material degradation function $\mathcal{D}(\mathfrak{g})$} Finally, exactly as the material degradation function (\ref{Dg-features}) introduced in the elementary plate problem, the function $\mathcal{D}(\mathfrak{g})$ in (\ref{Griffith-General}) stands for the material degradation function that describes the decrease of the critical energy release in the degradation regions $\Omega^{(s)}_\ell(t)$. It is a function that satisfies the properties
\begin{equation*}
0\leq \mathcal{D}\left(\mathfrak{g}\right)\leq 1,\quad \mathcal{D}\left(0\right)=1,\quad \mathcal{D}\left(\mathfrak{g}_1\right)\geq \mathcal{D}\left(\mathfrak{g}_2\right) \textrm{  for all   }
\mathfrak{g}_2>\mathfrak{g}_1,
\end{equation*}
but that otherwise it needs to be determined from experiments, much like the initial value $G_c$ of the critical energy release rate. 

\begin{remark}
\emph{Any experiment of choice that covers a sufficiently large range of values of the internal memory variable $\mathfrak{g}$ can be used to determine the material degradation function $\mathcal{D}\left(\mathfrak{g}\right)$. As illustrated in Subsection \ref{Sec: Dg-cali} above for the elementary plate problem, standard experiments from which Paris curves are generated (i.e., problems for specimens with simple geometries containing a centered or edge crack subjected to cyclic tensile loading) are a particularly appealing choice since $\mathfrak{g}$ can be computed explicitly for those and, at the same time, they typically cover a large range of values of $\mathfrak{g}$.}
\end{remark}

\begin{remark}
\emph{Exactly as the Young's modulus $E$, the Poisson's ratio $\nu$, and the initial value $G_c$ of the critical energy release rate, the material degradation function $\mathcal{D}\left(\mathfrak{g}\right)$ is an intrinsic macroscopic property of the material. As such, once it has been measured from an experiment for a material of interest, it is known once and for all for that material.}
\end{remark}

\subsection{Properties of the proposed Griffith formulation}\label{Sec: Formulation Properties}

In the sequel, we set forth the basic properties of the proposed Griffith formulation (\ref{Griffith-General}) and illustrate via examples how it describes the growth of large cracks in elastic brittle materials in a manner that is consistent with experimental observations, irrespective of how the loading is applied as a function of time. 

All the numerical examples that are presented below pertain to the same generic silicon nitride ceramic utilized in Sections \ref{Sec: Plate} and \ref{Sec: Griffith description} for the elementary plate problem. Tables \ref{Table1} and \ref{Table2} list the values of the elasticity constants and initial critical energy release rate for this representative material, as well as the corresponding constants in the formula (\ref{D-constitutive-formula}), used here to describe its material degradation function, for four different values of the characteristic size $\ell$ of the degradation region $\Omega_{\ell}(t)$.
\begin{table}[h!]\centering
\caption{Elasticity constants and initial critical energy release rate for silicon nitride.}
\begin{tabular}{cc|c}
\toprule
$E$ (GPa)& $\nu$  & $G_c$ (N/m) \\
\midrule
$300$ & $0.25$ & $120$ \\
\bottomrule
\end{tabular} \label{Table1}
\end{table}
\begin{table}[h!]\centering
\caption{Constants in the material degradation function (\ref{D-constitutive-formula}) for silicon nitride for four different values of the characteristic size $\ell$ of the degradation region $\Omega_{\ell}(t)$ .}
\begin{tabular}{l|ccccc}
\toprule
$\ell$ ($\mu$m) & $\dfrac{G_{\infty}}{G_c}$ & $\mathfrak{g}_{th}$ (N/m) & $K$ (m/N)$^{\alpha}$ & $\alpha$ & $\beta$\\
\midrule
$5$  & $0$ & $1439$ & $0.14$ & $1.2768$ & $1.1518$ \\
\midrule
$10$  & $0$ & $2874$ & $745$ & $0.2255$ & $0.1003$ \\
\midrule
$20$  & $0$ & $5734$ & $326$ & $0.2112$ & $0.0861$ \\
\midrule
$30$  & $0$ & $8579$ & $213$ & $0.1839$ & $ 0.0587$ \\
\bottomrule
\end{tabular} \label{Table2}
\end{table}

\subsubsection{Response under monotonic loading}

We begin by considering the limiting classical case when the loading is applied monotonically in time, that is, when 
\begin{equation*}
\overline{\bfu}(\bfX,t)=\overline{\bfu}_0(\bfX)t,\qquad\overline{\bfs}(\bfX,t)=\overline{\bfs}_0(\bfX)t,\qquad\textbf{b}(\bfX,t)=\textbf{b}_0(\bfX)t,
\end{equation*}
where $\overline{\bfu}_0(\bfX)$, $\overline{\bfs}_0$, and $\textbf{b}_0(\bfX)$ are suitably well-behaved functions of choice. 

Under this type of loading conditions, experiments indicate that large cracks grow in a manner that appears to be either continuous in time (stable crack growth) or discontinuous in time by brutal large jumps (unstable crack growth). In both scenarios, the internal memory variable (\ref{g-tau-Gen}) specializes to $\mathfrak{g}^{(s)}(t)=0$, the material degradation function remains then at its initial value $\mathcal{D}\left(\mathfrak{g}^{(s)}(t)\right)=\mathcal{D}\left(0\right)=1$, and consequently the critical energy release rate remains at its initial constant value $\mathcal{G}_c(\bfX,t)=G_c$ throughout the entire body for the entire duration $[0,T]$ of the loading process. As a result, for the case of monotonic loading conditions, the Griffith criticality condition (\ref{Griffith-General}) reduces identically to the classical Griffith criticality condition. 

\subsubsection{Response under cyclic loading with different constant maximum loads, load ratios, and frequencies}

\begin{figure}[H]
\centering
\centering\includegraphics[width=0.82\linewidth]{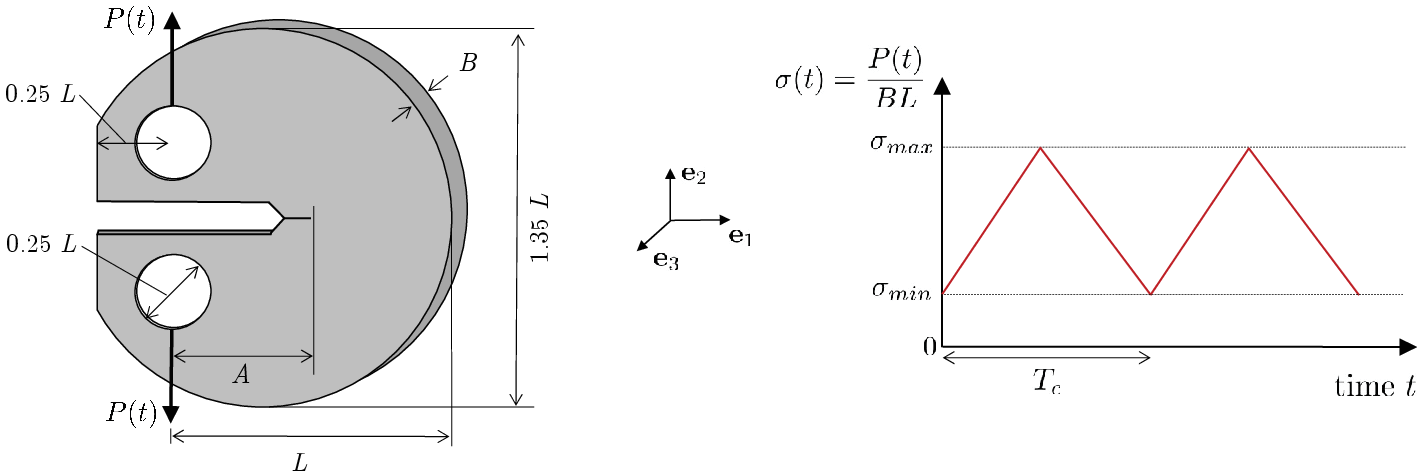}
\caption{\small Schematic of the disk-shaped compact tension test, containing a large pre-existing edge crack of initial size $A$, under linear cyclic loading, with cycle duration $T_c$, and hence loading frequency $f_c=1/T_c$, constant maximum stress $\sigma_{max}>0$, constant minimum stress $0\leq\sigma_{min}<\sigma_{max}$, and hence load ratio $R=\sigma_{min}/\sigma_{max}\geq 0$.}\label{Fig8}
\end{figure}

Next, we consider cyclic loadings with different constant maximum loads, load ratios, and frequencies. In particular, in view of its popular use by practitioners, we consider the disk-shaped compact tension test depicted in Fig. \ref{Fig8}  with applied linear cyclic force 
\begin{equation*}
P(t)=
\left\{\begin{array}{ll}
2\left[\dfrac{t}{T_c}-(N-1)\right]\left(1-\dfrac{P_{min}}{P_{max}}\right)P_{max}+P_{min}, & (N-1)T_c\leq t \leq (N-1)T_c+\dfrac{T_c}{2}\vspace{0.2cm}\\
2\left[-\dfrac{t}{T_c}+N\right]\left(1-\dfrac{P_{min}}{P_{max}}\right)P_{max}+P_{min}, & (N-1)T_c+\dfrac{T_c}{2}< t \leq N T_c
\end{array}\right. ,
\end{equation*}
where $P_{max}>P_{min}\geq 0$, $T_c$ stands for the cycle duration, and $N\in \mathbb{Z}^{+}$ denotes the number of loading cycles. Making use of the standard notation 
\begin{equation*}
f_c=\dfrac{1}{T_c},\qquad \sigma_{max}=\dfrac{P_{max}}{BL},\qquad \sigma_{min}=\dfrac{P_{min}}{BL},\qquad R=\dfrac{\sigma_{min}}{\sigma_{max}},
\end{equation*}
we parameterize the applied cyclic loading in terms of the global stress measure
\begin{equation}\label{eq_triangluar_load-CTT}
\sigma(t)=\dfrac{P(t)}{B L}=
\left\{\begin{array}{ll}
2\left[f_c t-(N-1)\right](1-R)\sigma_{max}+R\sigma_{max}, & (N-1)\leq f_c t \leq (N-1)+\dfrac{1}{2}\vspace{0.2cm}\\
2\left[-f_c t+N\right](1-R)\sigma_{max}+R\sigma_{max}, & (N-1)+\dfrac{1}{2}< f_c t \leq N 
\end{array}\right. .
\end{equation}

For this initial-boundary-value problem, wherein the crack is subject to Mode \texttt{I}, the stress intensity factor $K_{\texttt{I}}(t)$ is accurately described by the formula \citep{Tada73}
\begin{equation}\label{KI-CTT}
K_{\texttt{I}}(t)=\sigma(t)\sqrt{L}F(a(t))
\end{equation}
with
\begin{equation}\label{F(a)}
F(a(t))=\dfrac{\left(2+\dfrac{a(t)}{L}\right)}{\left(1-\dfrac{a(t)}{L}\right)^{3/2}}\left[0.76+4.8 \dfrac{a(t)}{L}-11.58\left(\dfrac{a(t)}{L}\right)^2+11.43 \left(\dfrac{a(t)}{L}\right)^3-4.08 \left(\dfrac{a(t)}{L}\right)^4\right],
\end{equation}
where $a(t)$ is the current size of the crack at time $t$, as measured from the axis where the force $P(t)$ is applied; see Fig. \ref{Fig8}. 

%
\begin{figure}[t!]
  \subfigure[]{
   \begin{minipage}[]{0.46\linewidth}
   \centering \includegraphics[width=\linewidth]{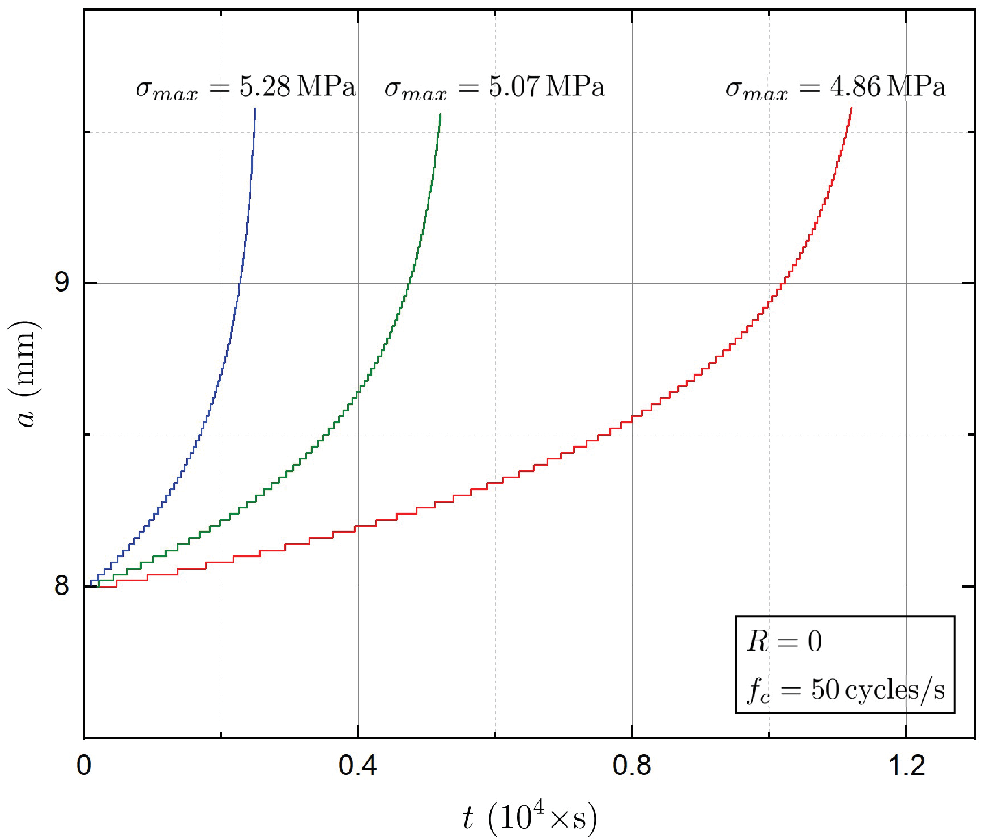}
   \end{minipage}}
  \subfigure[]{
   \begin{minipage}[]{0.485\linewidth}
   \centering \includegraphics[width=\linewidth]{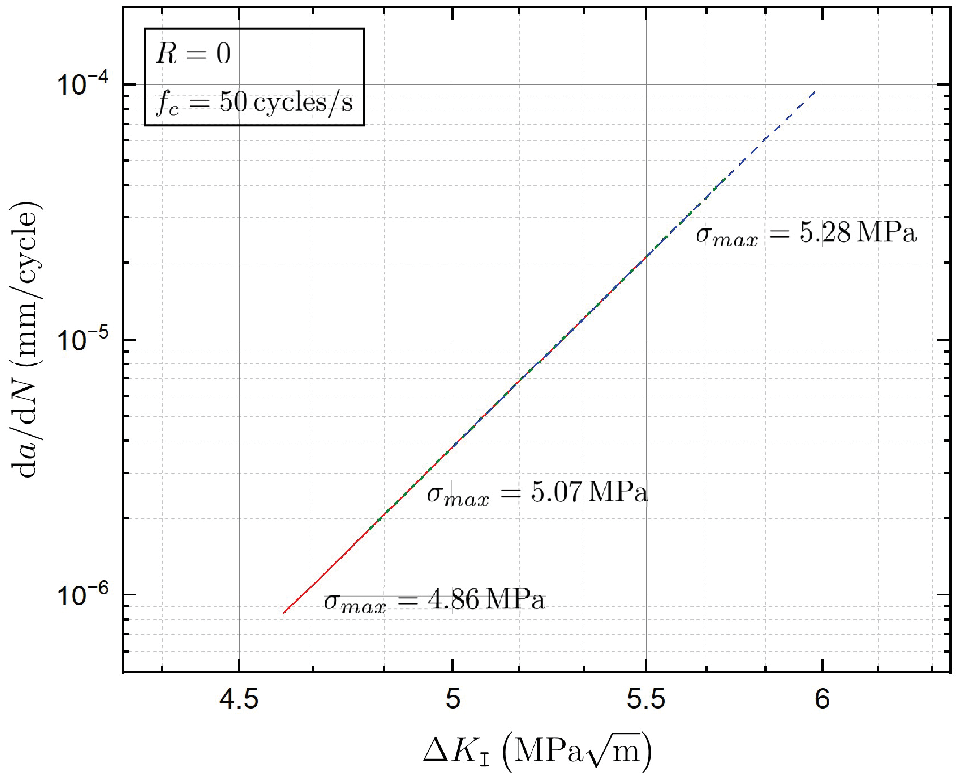}
   \end{minipage}}\par\centering
   \caption{Predictions generated by the Griffith criticality condition (\ref{Griffith-General-2D}) for the crack growth in disk-shaped compact tension tests under the linear cyclic loading (\ref{eq_triangluar_load-CTT}) with load ratio $R=0$, loading frequency $f_c=50$ cycles/s, and three different values of maximum stress, $\sigma_{max}= 4.86, 5.07, 5.28$ MPa. (a) The size $a(t)$ of the crack as a function of time. (b) The rate ${\rm d}a/{\rm d} N$  of crack growth as a function of the range of stress intensity factors $\Delta K_{\texttt{I}}$.}
   \label{Fig9}
\end{figure}
%

%
\begin{figure}[t!]
  \subfigure[]{
   \begin{minipage}[]{0.46\linewidth}
   \centering \includegraphics[width=\linewidth]{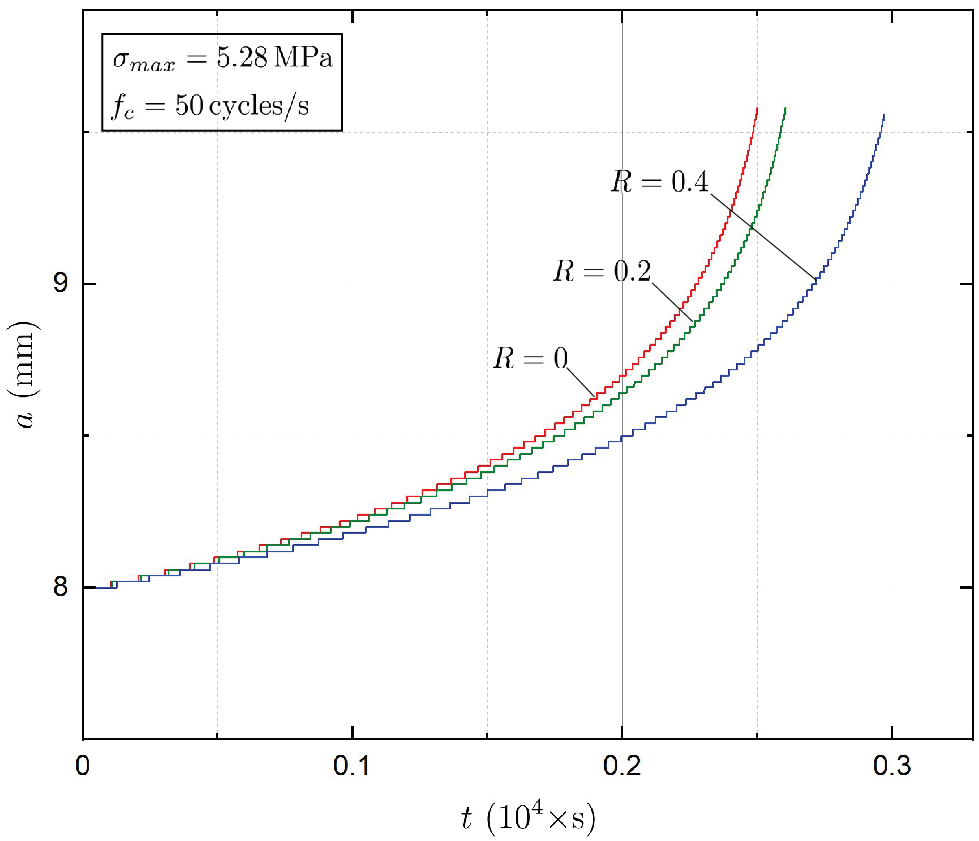}
   \end{minipage}}
  \subfigure[]{
   \begin{minipage}[]{0.485\linewidth}
   \centering \includegraphics[width=\linewidth]{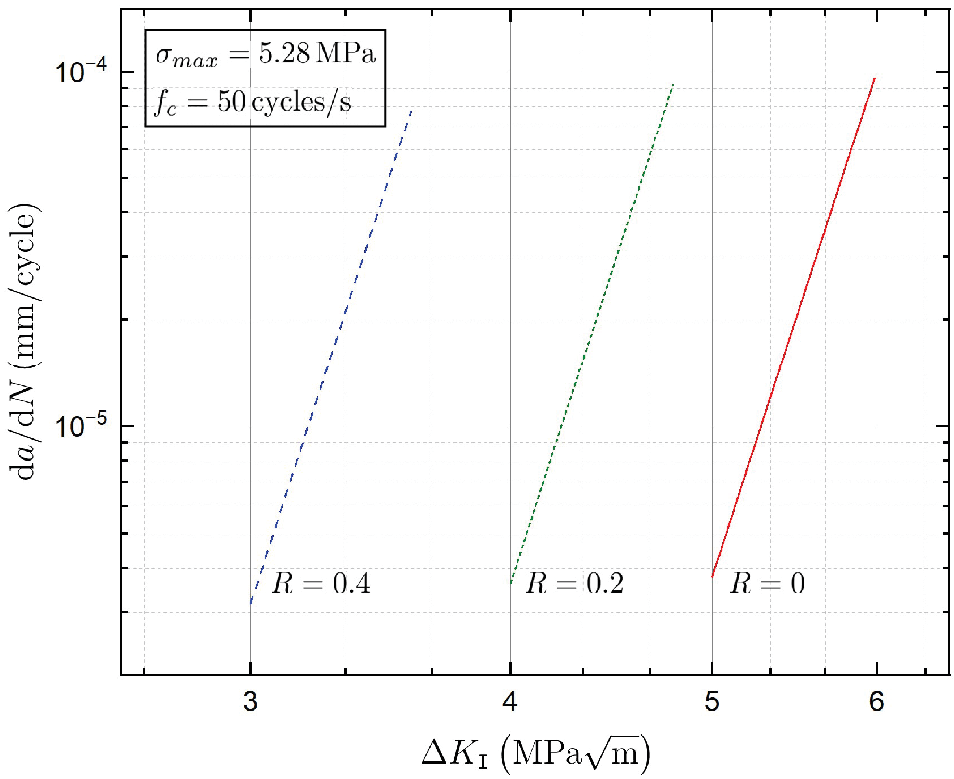}
   \end{minipage}}\par\centering
   \caption{Predictions generated by the Griffith criticality condition (\ref{Griffith-General-2D}) for the crack growth in disk-shaped compact tension tests under the linear cyclic loading (\ref{eq_triangluar_load-CTT}) with maximum stress $\sigma=5.28$ MPa, loading frequency $f_c=50$ cycles/s, and three different values of load ratio, $R= 0, 0.2, 0.4$. (a) The size $a(t)$ of the crack as a function of time. (b) The rate ${\rm d}a/{\rm d} N$  of crack growth as a function of the range of stress intensity factors $\Delta K_{\texttt{I}}$.}
   \label{Fig10}
\end{figure}
%

%
\begin{figure}[t!]
  \subfigure[]{
   \begin{minipage}[]{0.46\linewidth}
   \centering \includegraphics[width=\linewidth]{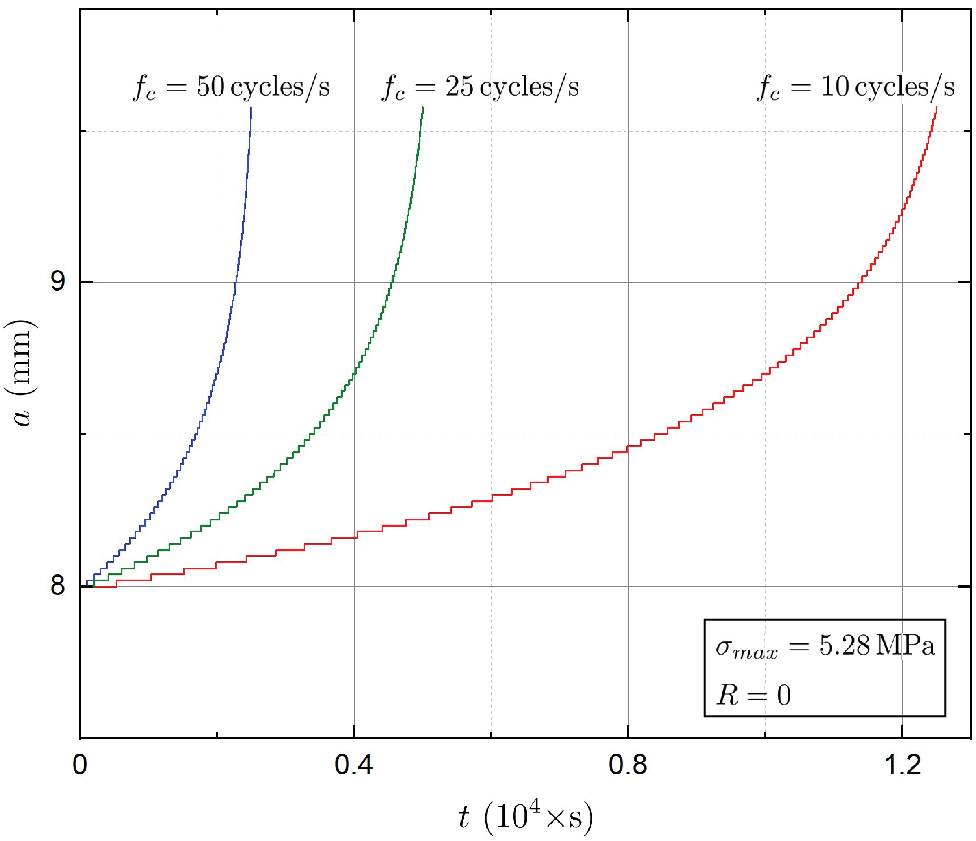}
   \end{minipage}}
  \subfigure[]{
   \begin{minipage}[]{0.485\linewidth}
   \centering \includegraphics[width=\linewidth]{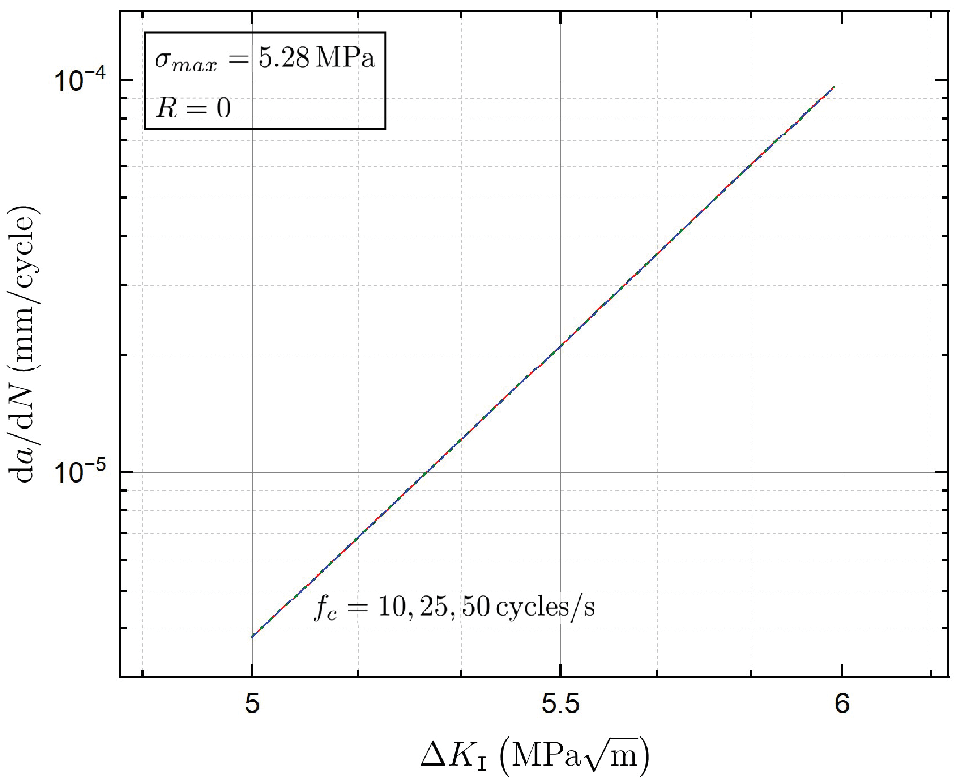}
   \end{minipage}}\par\centering
   \caption{Predictions generated by the Griffith criticality condition (\ref{Griffith-General-2D}) for the crack growth in disk-shaped compact tension tests under the linear cyclic loading (\ref{eq_triangluar_load-CTT}) with maximum stress $\sigma=5.28$ MPa, load ratio $R=0$, and three different values of loading frequency, $f_c= 10, 25, 50$ cycles/s. (a) The size $a(t)$ of the crack as a function of time. (b) The rate ${\rm d}a/{\rm d} N$  of crack growth as a function of the range of stress intensity factors $\Delta K_{\texttt{I}}$.}
   \label{Fig11}
\end{figure}
%

Making use of the formula (\ref{KI-CTT}) in expressions (\ref{ModeI-I1-W}) and, subsequently, making use of the resulting volume average quantities $\overline{I}_1(t)$ and $\overline{W}(t)$ in expression (\ref{g-tau-Gen}) yield an internal memory variable $\mathfrak{g}(t)$ that can be computed explicitly for any choice of the maximum stress $\sigma_{max}$, any choice of the load ratio $R$, and any choice of the loading frequency $f_c$ in the applied cyclic loading (\ref{eq_triangluar_load-CTT}). The result reads
\begin{align}\label{g-tau-RCT}
\mathfrak{g}(t)=\left\{\begin{array}{ll} 
0, & t_j=t \vspace{0.2cm}\\
\dfrac{k_0 L F^2(a(t_j))}{E}(1-R^2)\sigma^2_{max}\left(\left\lfloor f_c t \right\rfloor-\left\lfloor f_c t_j \right\rfloor\right) + c_l\left(t\right)- c_l\left(t_j\right), & t_j< t<t_{j+1} \end{array}\right. 
\end{align}
with
\begin{align*}
c_l\left(t\right)=\left\{\begin{array}{ll} 
\dfrac{k_0 L F^2(a(t_j))}{E}\sigma^2_{max}\left(2 (1-R) \left(\left\lfloor\dfrac{t}{T_c}\right\rfloor-\dfrac{t}{T_c}\right)-R\right)^2, & \left\lfloor\dfrac{t}{T_c}\right\rfloor T_c\leq t\leq \left\lfloor\dfrac{t}{T_c}\right\rfloor T_c + \dfrac{T_c}{2} \vspace{0.2cm}\\
\dfrac{k_0 L F^2(a(t_j))}{E}\sigma^2_{max}, & \left\lfloor\dfrac{t}{T_c}\right\rfloor T_c + \dfrac{T_c}{2}\leq t\leq \left(\left\lfloor\dfrac{t}{T_c}\right\rfloor +1\right) T_c  \end{array}\right. .
\end{align*}
In turn, the material degradation function $\mathcal{D}\left(\mathfrak{g}(t)\right)$ and hence the evolution of the critical energy release rate $\mathcal{G}_c(\bfX,t)$ can be readily determined along the entire loading process for any choice of $\sigma_{max}$, $R$, and $f_c$. As we illustrate next via numerical examples, this evolution is such that the Griffith criticality condition (\ref{Griffith-General-2D}) predicts that the crack grows by jumps of length scale $\ell$, in a manner that, consistent with experimental observations, depends on all three loading parameters $\sigma_{max}$, $R$, and $f_c$.

Figures \ref{Fig9}, \ref{Fig10}, and \ref{Fig11} present results for the crack growth in disk-shaped compact tension tests, under the linear cyclic loading (\ref{eq_triangluar_load-CTT}), as predicted by the Griffith criticality condition (\ref{Griffith-General-2D}). The results pertain to specimens of length $L=24.9$ mm, thickness $B=1.9$ mm, and initial crack size $A=8$ mm, that are made of a silicon nitride with the material constants listed in Table \ref{Table1} and those listed in Table \ref{Table2} for the characteristic size $\ell=20$ $\mu$m of the degradation region $\Omega_\ell(t)$. Parts (a) of the figures provide plots for the size $a(t)$ of the crack as a function of time $t$, while parts (b) provide plots for the rate ${\rm d}a/{\rm d} N$ of crack growth as a function of the range of stress intensity factors $\Delta K_{\texttt{I}}=(1-R)\sigma_{max}\sqrt{L}F(a)$. Specifically, on the one hand, Fig. \ref{Fig9} shows results for the fixed load ratio $R=0$ and three different values of maximum stress, $\sigma_{max}= 4.86, 5.07, 5.28$ MPa, while Fig. \ref{Fig10} shows results for the fixed maximum stress $\sigma_{max}=5.28$ MPa and three different values of load ratio, $R= 0, 0.2, 0.4$. All the results in Figs. \ref{Fig9} and \ref{Fig10} correspond to the same loading frequency $f_c=50$ cycles/s. On the other hand,  Fig. \ref{Fig11} shows results for fixed maximum stress $\sigma_{max}=5.28$ MPa, fixed load ratio $R=0$, and three different values of loading frequency, $f_c=10, 25, 50$ cycles/s.

The main observation from Figs. \ref{Fig9}(a), \ref{Fig10}(a), and \ref{Fig11}(a) is that the crack growth in disk-shaped compact tension tests is indeed significantly affected by the maximum stress $\sigma_{max}$, the load ratio $R$, and the loading frequency $f_c$ in the applied cyclic loading. Specifically, as expected from basic physical intuition, larger values of the maximum stress $\sigma_{max}$ and of the loading frequency $f_c$ lead to faster growths of the crack, while larger load ratios $R$ lead to slower growths. From Figs. \ref{Fig9}(b), \ref{Fig10}(b), and \ref{Fig11}(b), moreover, the main observation is that the corresponding growth rates ${\rm d}a/{\rm d} N$ show the typical Paris-law behavior observed in experiments. What is more, Fig. \ref{Fig10}(b) shows that larger values of the load ratio $R$ result in a left translation of the Paris curves. On the other hand, Figs. \ref{Fig9}(b) and \ref{Fig11}(b) show that larger values of the maximum stress $\sigma_{max}$ result in a shift along the same Paris curve, while different values the loading frequency $f_c$ have no impact on the Paris curves.

\subsubsection{Response under cyclic loading with nonlinear waveforms and overloads}

%
\begin{figure}[b!]
\centering \includegraphics[scale=0.48]{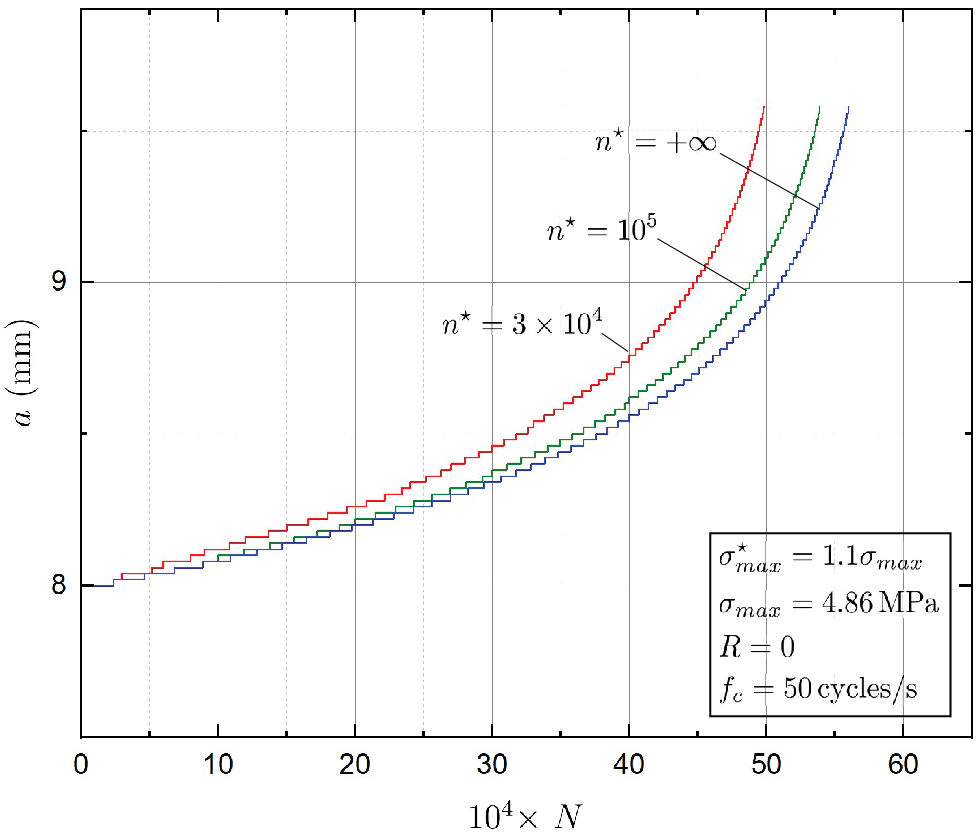}
\caption{Predictions generated by the Griffith criticality condition (\ref{Griffith-General-2D}) for the growth of the size $a(t)$ of the crack in disk-shaped compact tension tests under the sinusoidal cyclic loading (\ref{eq_sin_load-CTT}), with an overload maximum stress $\sigma^\star_{max}$ applied every $n^\star=3\times10^4, 10^5,$ and $+\infty$ loading cycles, plotted as a function of the number $N$ of loading cycles.}
\label{Fig12}
\end{figure}
%

We proceed by considering cyclic loadings with a nonlinear waveform --- as opposed to the linear waveform used in (\ref{eq_triangluar_load-CTT}) --- that may also contain overloads, this still within the initial-boundary-value problem of the disk-shaped compact tension test. In particular, for definiteness, we consider that the specimen is subjected to the sinusoidal cyclic global stress
\begin{equation}\label{eq_sin_load-CTT}
\sigma(t)=\dfrac{P(t)}{B L}=\left\{\begin{array}{ll}\dfrac{1+R^{\star}}{2}\sigma^{\star}_{max}+\dfrac{1-R^{\star}}{2}\sigma^{\star}_{max}\sin\left(2\pi f_c t-\dfrac{\pi}{2}\right), &  N-1\leq f_c t \leq N\quad {\rm and}\quad \dfrac{N}{n^\star}\in \mathbb{Z}^+\vspace{0.2cm}\\
\dfrac{1+R}{2}\sigma_{max}+\dfrac{1-R}{2}\sigma_{max}\sin\left(2\pi f_c t-\dfrac{\pi}{2}\right), & {\rm else}\end{array} \right.,
\end{equation}
in which an overload $P^\star_{max}\geq P_{max}$ is applied every $n^\star\in\mathbb{Z}^{+}$ loading cycles and where use has been made of the notation $\sigma^{\star}_{max}=P^\star_{max}/(B L)$ and $R^\star=\sigma_{min}/\sigma^{\star}_{max}$.

Similar to the linear cyclic global stress (\ref{eq_triangluar_load-CTT}) considered above, the internal memory variable $\mathfrak{g}(t)$ that results from the sinusoidal global stress (\ref{eq_sin_load-CTT}) can be computed explicitly. Precisely, substitution of (\ref{eq_sin_load-CTT})  in the formula (\ref{KI-CTT}) for the stress intensity factor and the successive evaluations of the average volume quantities (\ref{ModeI-I1-W}) and of the integral in (\ref{g-tau-Gen}) lead to the internal memory variable 
\begin{align}\label{gt-RCT-Overload}
\mathfrak{g}(t)=\left\{\begin{array}{ll} 
0, & t_j=t \vspace{0.2cm}\\
\dfrac{k_0 L F^2(a(t_j))}{E}\left[\dfrac{(n^\star-1)(1-R^2)}{n^\star}\sigma^2_{max}+\dfrac{1}{n^\star}(1-{R^{\star}}^2){\sigma^\star}^2_{max}\right]\left(\left\lfloor f_c t\right\rfloor-\left\lfloor f_c t_j \right\rfloor\right) + c_s(t)-c_s(t_j), & t_j< t<t_{j+1} \end{array}\right. ,
\end{align}
where $c_s(t)$ is a correction term, different from the correction $c_l(t)$ in (\ref{g-tau-RCT}), not spelled out here for conciseness. With the result (\ref{gt-RCT-Overload}) at out disposal, the material degradation function $\mathcal{D}\left(\mathfrak{g}(t)\right)$ and hence the evolution of the critical energy release rate $\mathcal{G}_c(\bfX,t)$ can be readily determined along the entire loading process for any choice of the maximum stress $\sigma_{max}$, the load ratio $R$, the loading frequency $f_c$, the overload maximum stress $\sigma^\star_{max}$, and the number $\left\lfloor N/n^\star \right\rfloor$ of instances at which the overload is applied. As in the preceding subsection, this evolution is such that the Griffith criticality condition (\ref{Griffith-General-2D}) predicts that the crack grows by jumps of length scale $\ell$. 

In the absence of an overload, when $\sigma^\star_{max}=\sigma_{max}$, we begin by noting that, up to the difference in the correction terms $c_l(t)$ and $c_s(t)$, both of which are typically small, the internal memory variable (\ref{gt-RCT-Overload}) is identical to the internal memory variable  (\ref{g-tau-RCT}) for the case when the cyclic load is applied linearly. This implies that the use of different waveforms for the applied cyclic load in disk-shaped compact tension tests has little impact on the growth of the crack. This prediction by the Griffith criticality condition (\ref{Griffith-General-2D}) is yet again consistent with experimental observations; see, e.g., \cite{Ritchie91}.

In contrast to the type of waveform, the presence of overloads does have a significant effect on how cracks grow in disk-shaped compact tension tests. To illustrate their impact, Fig. \ref{Fig12} presents results for the crack growth in disk-shaped compact tension tests, as predicted by the Griffith criticality condition (\ref{Griffith-General-2D}), under the sinusoidal cyclic loading (\ref{eq_sin_load-CTT}) with $\sigma_{max}=4.86$ MPa, $R=0$, $f_c=50$ cycles/s, $\sigma^\star_{max}=1.1\sigma_{max}=5.35$ MPa, for three different values of the number of cycles at which the overload is applied, $n^\star=3\times10^4, 10^5,$ and $+\infty$. The results, which are shown in terms of the size $a(t)$ of the crack as a function of the number of loading cycles $N=f_c t$, pertain to the same specimens used above, that is, specimens of length $L=24.9$ mm, thickness $B=1.9$ mm, and initial crack size $A=8$ mm, that are made of a silicon nitride with the material constants listed in Table \ref{Table1} and those listed in Table \ref{Table2} for the characteristic size $\ell=20$ $\mu$m of the degradation region $\Omega_\ell(t)$. 

As expected from basic physical intuition, the results in Fig. \ref{Fig12} show that the presence of overloads accelerates the growth of the crack, more so the larger the number of instances at which the overload is applied.

\subsubsection{Sensitivity to the choice of the characteristic size $\ell$ of the degradation region $\Omega_\ell(t)$}\label{Sec: ell sensitivity}

%
\begin{figure}[b!]
  \subfigure[]{
   \begin{minipage}[]{0.46\linewidth}
   \centering \includegraphics[width=\linewidth]{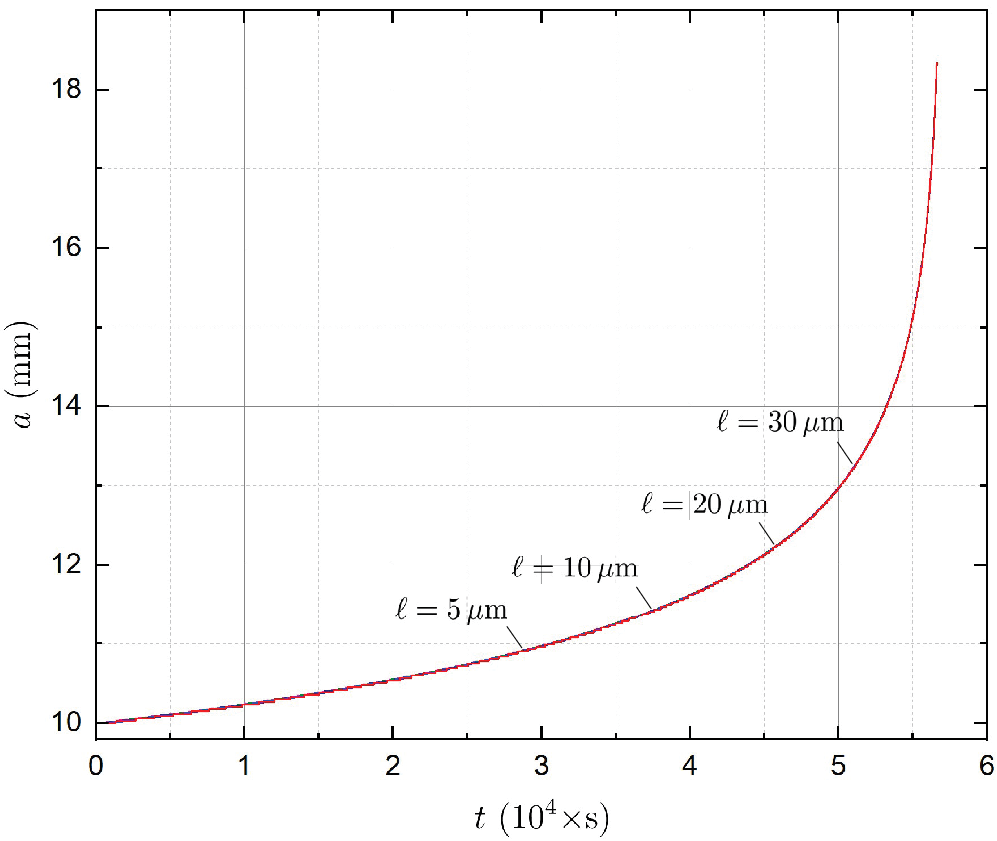}
   \end{minipage}}
  \subfigure[]{
   \begin{minipage}[]{0.48\linewidth}
   \centering \includegraphics[width=\linewidth]{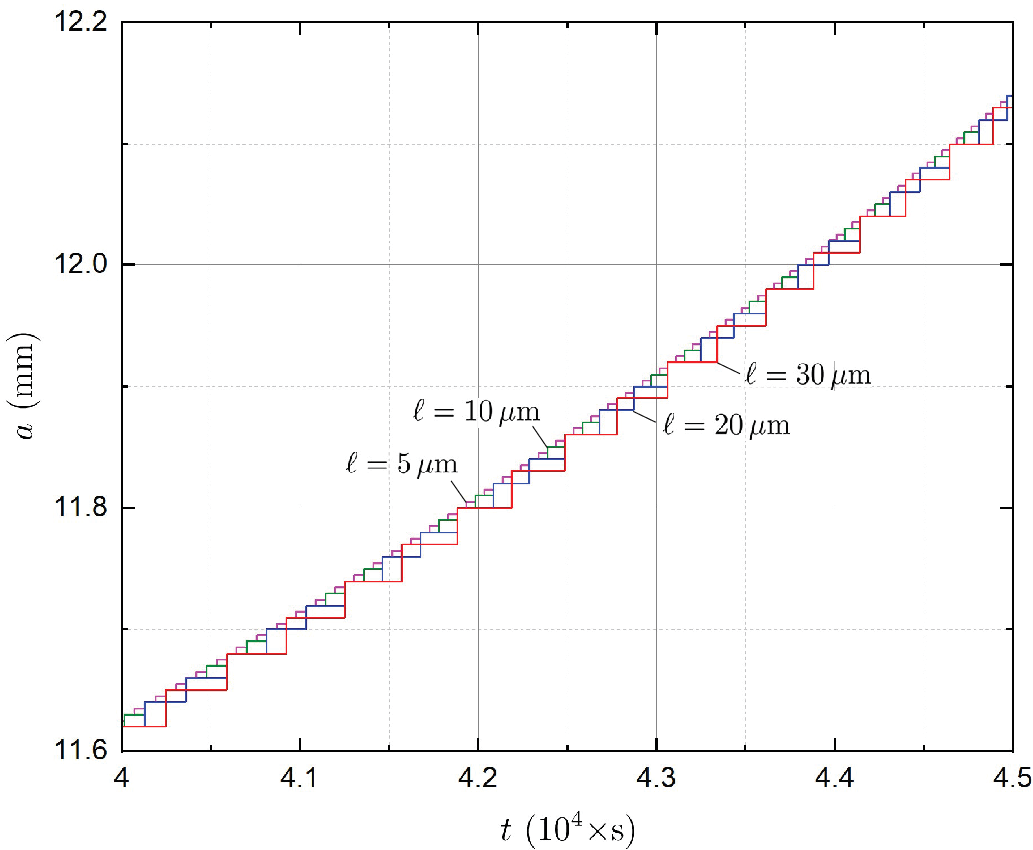}
   \end{minipage}}\par\centering
   \caption{Time-discontinuous evolution of the half size (\ref{crack_evolution}) of a crack, of initial half size $A=10$ mm, by jumps of  constant sizes $\ell=5, 10, 20, 30$ $\mu$m. The results pertain to same initial-boundary-value problem considered in Fig. \ref{Fig3} for a silicon nitride. (a) Plots of $a(t)$  over the entire loading interval $[0,5.6615\times10^4]$ s. (b)  Close-up over the interval $[4.00\times 10^4,4.50\times 10^4]$ s.}
   \label{Fig13}
\end{figure}
%

In the extraction of the material degradation function $\mathcal{D}(\mathfrak{g})$ from the measured growth of the size $a(t)$ of a crack in an experiment, one must choose a specific value for the characteristic size $\ell$ of the degradation region $\Omega_\ell(t)$. Naturally, different choices of $\ell$ lead to different material degradation functions $\mathcal{D}(\mathfrak{g})$. In this subsection, we show via an example that the Griffith criticality condition (\ref{Griffith-General}) is largely insensitive to the specific choice of characteristic size $\ell$, so long as the material degradation function $\mathcal{D}(\mathfrak{g})$ utilized in (\ref{Griffith-General}) corresponds to the chosen $\ell$ and so long as the chosen $\ell$ does not exceed the correct order of magnitude. 

%
\begin{figure}[t!]
  \subfigure[]{
   \begin{minipage}[]{0.48\linewidth}
   \centering \includegraphics[width=\linewidth]{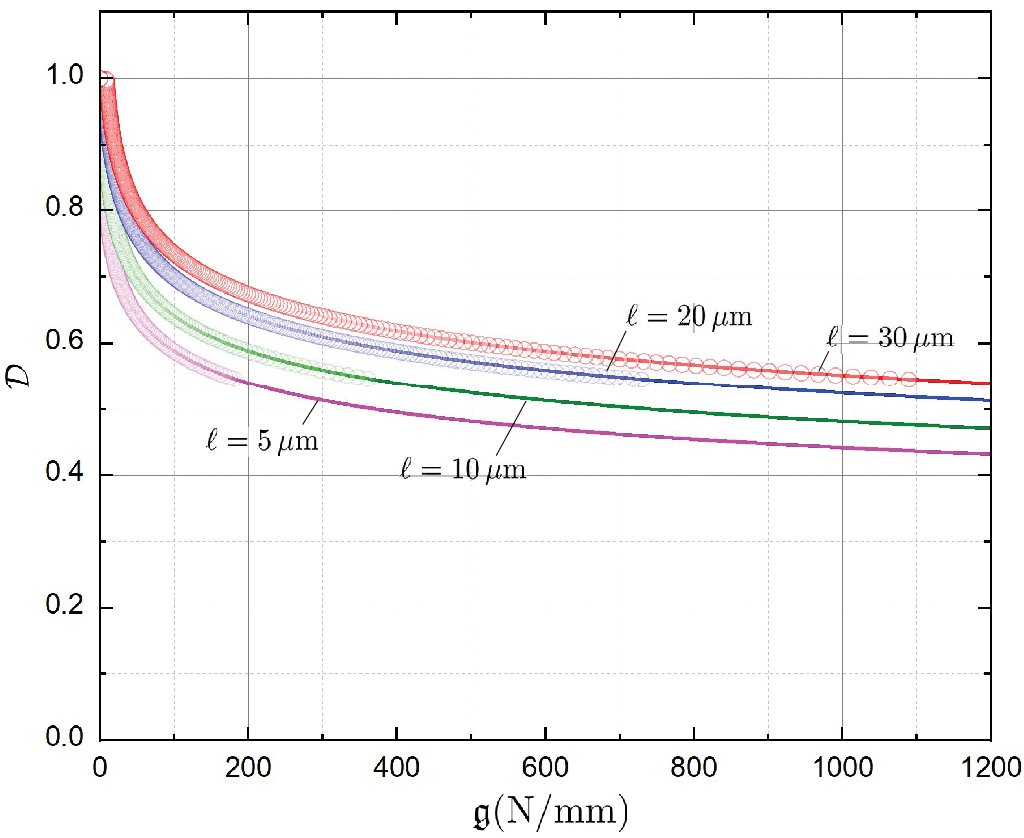}
   \end{minipage}}
  \subfigure[]{
   \begin{minipage}[]{0.47\linewidth}
   \centering \includegraphics[width=\linewidth]{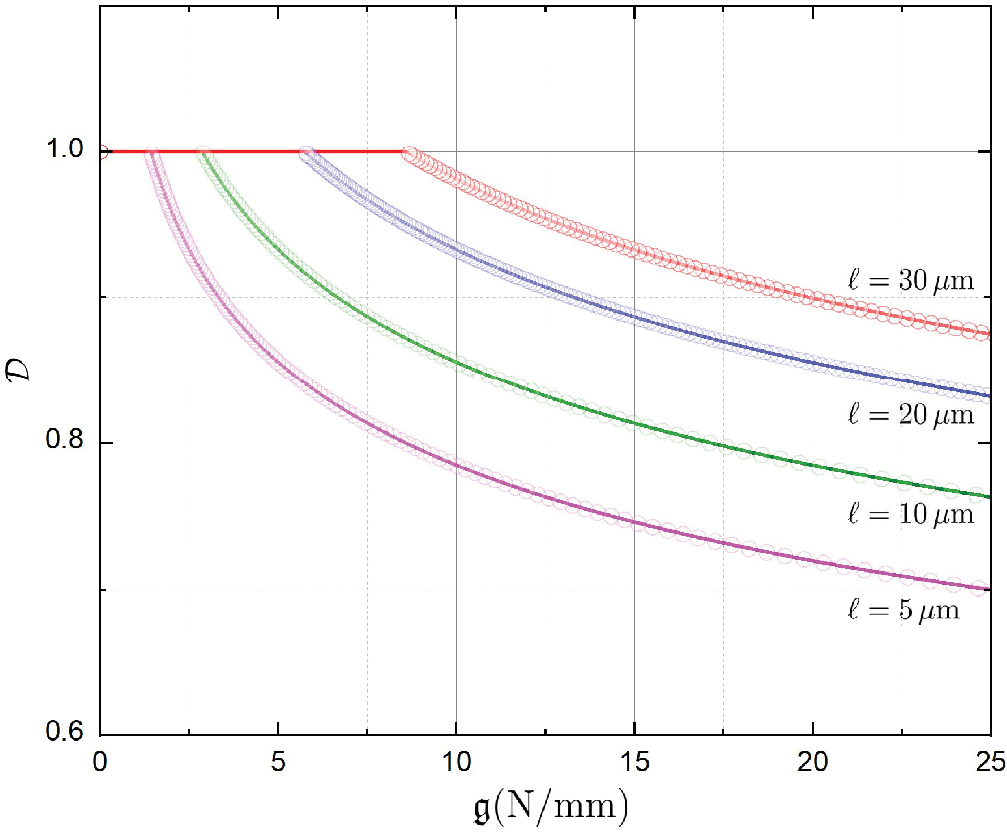}
   \end{minipage}}\par\centering
   \caption{The material degradation functions $\mathcal{D}(\mathfrak{g})$ extracted from the crack growths in Fig. \ref{Fig13} for four different choices of the characteristic size $\ell$ of the degradation region $\Omega_\ell(t)$. (a) Plot of $\mathcal{D}(\mathfrak{g})$  over the entire range of $\mathfrak{g}$ values. (b)  Close-up around $\mathfrak{g}=\mathfrak{g}_{th}$.}
   \label{Fig14}
\end{figure}
%
%
\begin{figure}[t!]
  \subfigure[]{
   \begin{minipage}[]{0.46\linewidth}
   \centering \includegraphics[width=\linewidth]{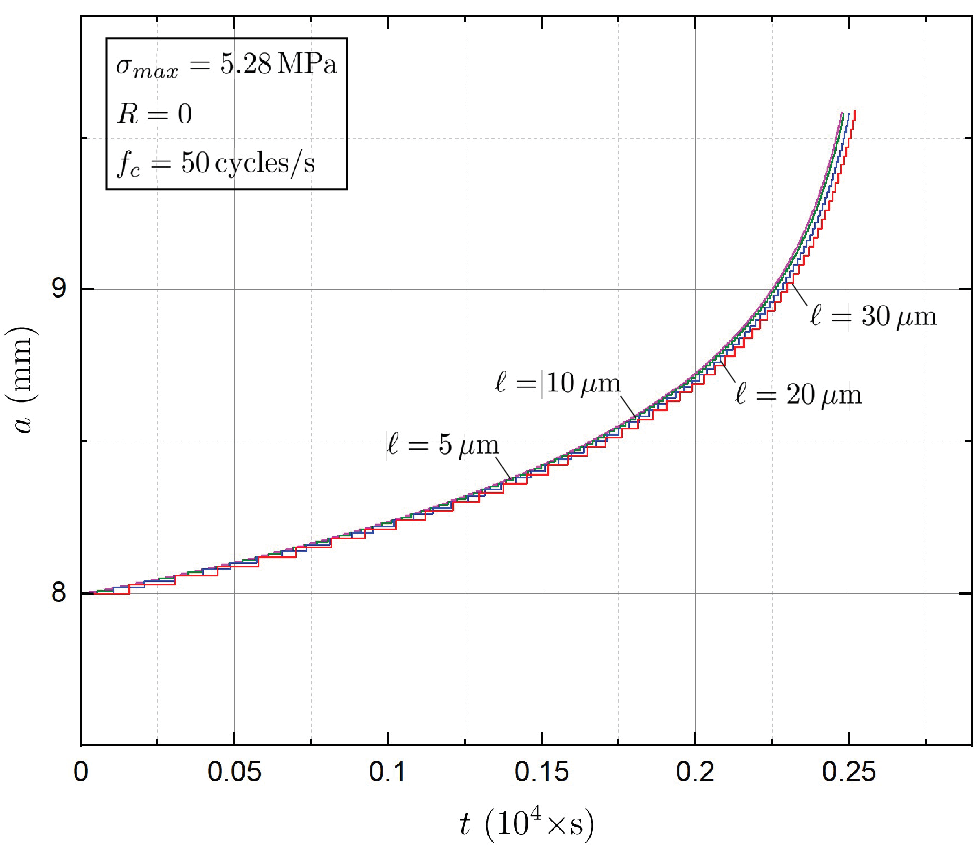}
   \end{minipage}}
  \subfigure[]{
   \begin{minipage}[]{0.485\linewidth}
   \centering \includegraphics[width=\linewidth]{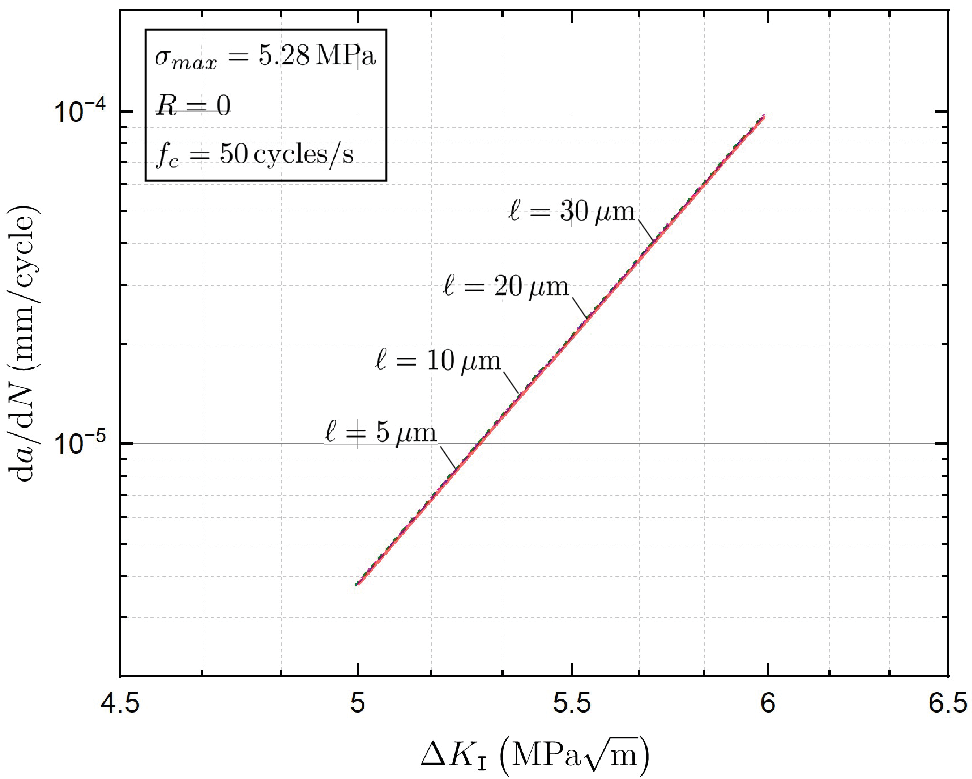}
   \end{minipage}}\par\centering
   \caption{Predictions generated by the Griffith criticality condition (\ref{Griffith-General-2D}) for the crack growth in a disk-shaped compact tension test under the linear cyclic loading (\ref{eq_triangluar_load-CTT}) with maximum stress $\sigma=5.28$ MPa, load ratio $R=0$, and loading frequency $f_c=50$ cycles/s, for four different values of the characteristic size of the degradation region $\Omega_\ell(t)$, $\ell=5, 10,20, 30$ $\mu$m. (a) The size $a(t)$ of the crack as a function of time. (b) The rate ${\rm d}a/{\rm d} N$  of crack growth as a function of the range of stress intensity factors $\Delta K_{\texttt{I}}$.}
   \label{Fig15}
\end{figure}
%

Consider again the growth of the half size $a(t)$ of the crack presented in Fig. \ref{Fig3} for the elementary plate problem analyzed in Sections \ref{Sec: Plate} and \ref{Sec: Griffith description}. That crack growth corresponds to the choice of $\ell=20$ $\mu$m. Figure \ref{Fig13} presents plots of the same crack growth for $\ell=20$ $\mu$m and for the three additional values $\ell=5, 10,$ and $\ell=30$ $\mu$m. Use of these four sets of crack-growth descriptions in equations (\ref{g-tau-Plate}), (\ref{Dg-features})$_2$, and (\ref{Dg-features})$_3$ leads to the four different material degradation functions $\mathcal{D}(\mathfrak{g})$ plotted (circles) in Fig. \ref{Fig14}. As expected, all four are different. In spite of their differences, all four material degradation functions can be accurately approximated by the formula (\ref{D-constitutive-formula}) with the constants listed in Table \ref{Table2} for each $\ell$. For direct comparison, Fig. \ref{Fig14} includes plots (solid lines) of this formula for the four characteristic sizes $\ell=5, 10, 20, 30$ $\mu$m.

Armed with the four material degradation functions (\ref{D-constitutive-formula}), with the constants listed in Table \ref{Table2} for the four characteristic sizes $\ell=5, 10, 20, 30$ $\mu$m, we can now deploy the Griffith criticality condition (\ref{Griffith-General}) to investigate how the choice of the value of $\ell$ impacts its predictions for crack growth in the generic silicon nitride used throughout this section. Here, we do so again for the experimentally prominent disk-shaped compact tension test. Specifically, we consider the same specimen used above with length $L=24.9$ mm, thickness $B=1.9$ mm, and initial crack size $A=8$ mm, under the linear cyclic loading (\ref{eq_triangluar_load-CTT})  with maximum stress $\sigma=5.28$ MPa, load ratio $R=0$, and loading frequency $f_c=50$ cycles/s. Figures \ref{Fig15}(a) and \ref{Fig15}(b) present the results generated by the Griffith criticality condition (\ref{Griffith-General-2D}) for the size $a(t)$ of the crack as a function of time, as well as for the corresponding rate ${\rm d}a/{\rm d}N$ of crack growth as a function of the range of stress intensity factor $\Delta K_{\texttt{I}}$.

A quick glance at the results in Fig. \ref{Fig15} suffices to recognize that the Griffith criticality condition (\ref{Griffith-General-2D}) is indeed largely insensitive to the specific choice of the characteristic size $\ell$ of the degradation region $\Omega_\ell(t)$. The results further suggest that there is a limit response as $\ell\searrow 0$ that may be equivalent to the one associated with the actual value of $\ell$. We shall explore this limit in future work.

\section{A first set of validation results}\label{Sec: Validation}

The examples presented in the preceding section have shown that the proposed Griffith criticality condition (\ref{Griffith-General}) describes the growth of large cracks in elastic brittle materials in a manner that is \emph{qualitatively} consistent with experimental observations, irrespective of how the loading is applied as a function of time. In this section, we show the same in a \emph{quantitative} manner by means of comparisons with three sets of fatigue fracture experiments on a commercial-grade silicon nitride ceramic \citep{Ritchie95}, on mortar \citep{Wang91}, and on a commercial PMMA \citep{Clark90}.

\subsection{Comparisons with the experiments of \cite{Ritchie95} on a commercial silicon nitride}

\cite{Ritchie95} carried out disk-shaped compact tension tests (see Fig. \ref{Fig8}) on specimens of length $L=24.9$ mm, thickness $B=1.9$ mm, and initial crack size $A>5$ mm that were made of the commercial silicon nitride EC-141 manufactured by NTK Technical Ceramic. Table \ref{Table3} lists the values of the elasticity constants and initial critical energy release rate for this material, as reported by \cite{Ritchie95}. These authors also reported that the average grain size was about 2 $\mu$m. The specimens were subjected to the sinusoidal loading (\ref{eq_sin_load-CTT}) with load ratios $R=0.1, 0.5, 0.7$,  no overload $\sigma^\star_{max}=\sigma_{max}$, and loading frequency $f_c=25$ cycles/s. While \cite{Ritchie95} did not report the precise initial size $A$ of the crack or the precise value of the maximum stress $\sigma_{max}$ that they applied, the choices $A=8$ mm and $\sigma_{max}=4$ MPa appear to be consistent with all their data. We therefore consider that $A=8$ mm and $\sigma_{max}=4$ MPa in their experiments. The main experimental results presented by \cite{Ritchie95} correspond to the three Paris curves that they plotted in Fig. 6 of their work for each of the three load ratios $R=0.1, 0.5, 0.7$ that they considered. 

\begin{table}[H]\centering
\caption{Elasticity constants and initial critical energy release rate for the commercial silicon nitride EC-141.}
\begin{tabular}{cc|c}
\toprule
$E$ (GPa)& $\nu$  & $G_c$ (N/m) \\
\midrule
$310$ & $0.27$ & $108$ \\
\bottomrule
\vspace{-0.2cm}
\end{tabular} \label{Table3}
\end{table}
\begin{table}[H]\centering
\caption{The characteristic size $\ell$ of the degradation region $\Omega_{\ell}(t)$ and the associated constants in the material degradation function (\ref{D-constitutive-formula}) for the commercial silicon nitride EC-141.}
\begin{tabular}{l|ccccc}
\toprule
$\ell$ ($\mu$m) & $\dfrac{G_{\infty}}{G_c}$ & $\mathfrak{g}_{th}$ (N/m) & $K$ (m/N)$^{\alpha}$ & $\alpha$ & $\beta$\\
\midrule
$20$  & $0$ & $16$ & $12440$ & $0.8571$ & $0.7847$ \\
\bottomrule
\end{tabular} \label{Table4}
\end{table}
%

%
\begin{figure}[H]
  \subfigure[]{
   \begin{minipage}[]{0.47\linewidth}
   \centering \includegraphics[width=\linewidth]{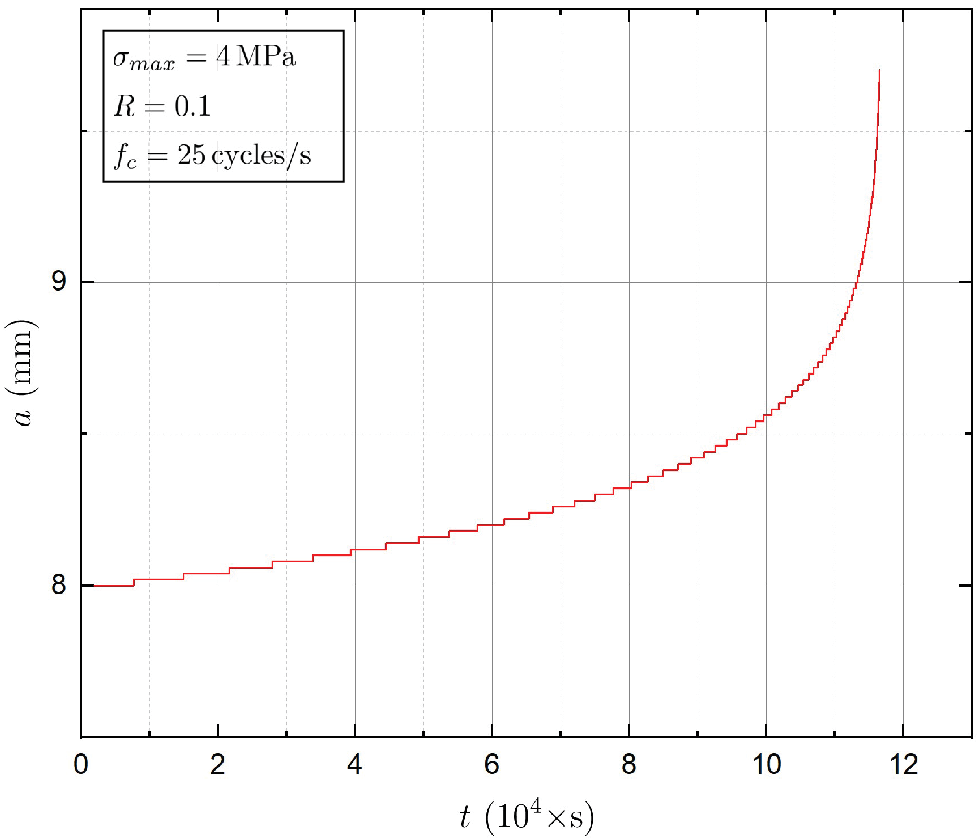}
   \end{minipage}}
  \subfigure[]{
   \begin{minipage}[]{0.48\linewidth}
   \centering \includegraphics[width=\linewidth]{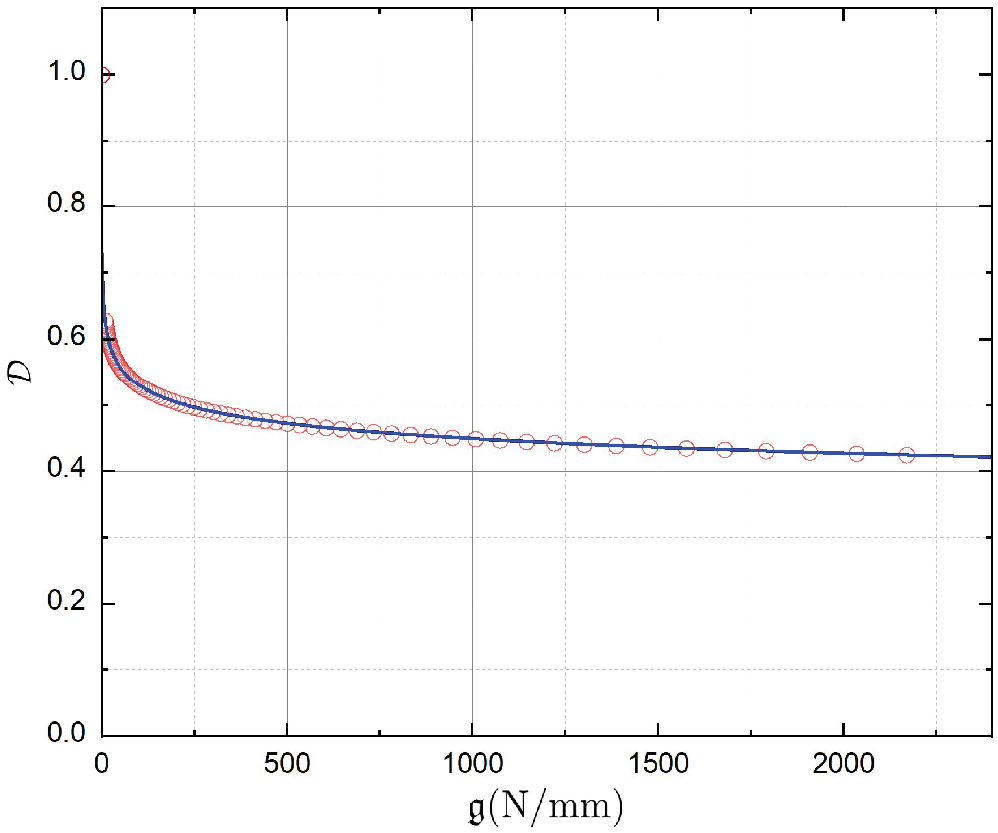}
   \end{minipage}}\par\centering
   \caption{(a) The size $a(t)$ of the crack as a function of time from the disk-shaped compact tension test of \cite{Ritchie95} on the commercial silicon nitride EC-141 for the load ratio $R=0.1$. The result corresponds to the choice of $\ell=20$ $\mu$m for the characteristic size of the degradation region $\Omega_{\ell}(t)$. (b) Plot (circles) of the corresponding degradation function $\mathcal{D}(\mathfrak{g})$. For direct comparison, the formula (\ref{D-constitutive-formula}), with the material constants listed in Table \ref{Table4}, is also plotted (solid line).}
   \label{Fig16}
\end{figure}
%

In order to confront the predictions generated by the Griffith criticality condition (\ref{Griffith-General-2D}) to the experiments of \cite{Ritchie95}, we must first determine an appropriate value for the characteristic size $\ell$ of the degradation region $\Omega_{\ell}(t)$, as well as the associated material degradation function $\mathcal{D}(\mathfrak{g})$ for the silicon nitride EC-141 that they tested. 

Given that the underlying microstructure of the material featured an average grain size of about 2 $\mu$m, we set $\ell=20$ $\mu$m. 

Granted the characteristic size $\ell=20$ $\mu$m, in order to determine the material degradation function $\mathcal{D}(\mathfrak{g})$, we proceed by integrating one of the Paris curves reported by \cite{Ritchie95}, say that for load ratio $R=0.1$, so as to have a set of experimental data for the growth of the size $a(t)$ of a crack under cyclic loading. Then, making use of this data for $a(t)$ and the result (\ref{gt-RCT-Overload}) for the internal memory variable $\mathfrak{g}(t)$ relevant to this problem, direct use of the relations 
\begin{equation*}
0\leq \mathcal{D}\left(\mathfrak{g}\right)\leq 1,\quad \mathcal{D}\left(0\right)=1,\quad \displaystyle\lim_{t\nearrow t_{j+1}}\mathcal{D}\left(\mathfrak{g}(t)\right)=\dfrac{\left(\sigma(t_{j+1})\sqrt{L}F(a_j)\right)^2}{G_c E},
\end{equation*}%
where we recall that for a disk-shaped compact tension test the function $F(a(t))$ is given by (\ref{F(a)}), allows to construct the material degradation function $\mathcal{D}(\mathfrak{g})$. Figure \ref{Fig16}(a) plots the size $a(t)$ of the crack as a function of time for the load ratio $R=0.1$, while Fig. \ref{Fig16}(b) plots (circles) the corresponding material degradation function $\mathcal{D}(\mathfrak{g})$. 

As has been the case in the preceding sections, the material degradation function $\mathcal{D}(\mathfrak{g})$ can be accurately described by the formula (\ref{D-constitutive-formula}), in this case, with the constants listed in Table \ref{Table4}. For direct comparison, Fig. \ref{Fig16}(b) includes the plot (solid line) of this formula with such constants. 

\begin{figure}[t!]
\centering \includegraphics[scale=0.5]{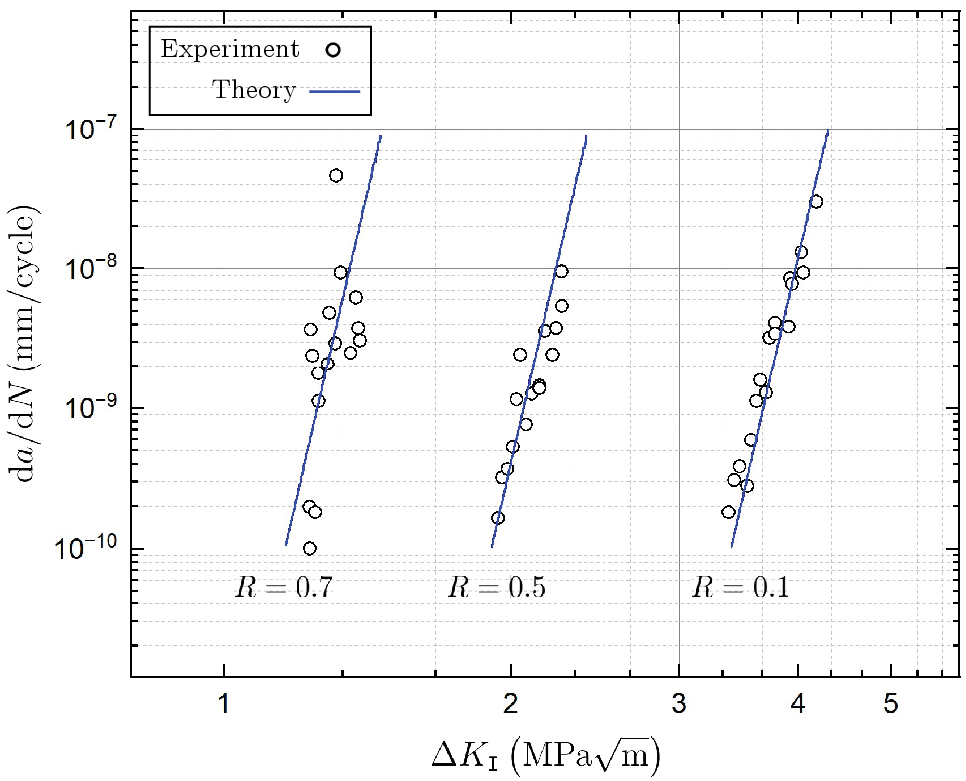}
\caption{Comparison between the predictions (solid lines) generated by the Griffith criticality condition (\ref{Griffith-General-2D}) and the experiments (circles) of \cite{Ritchie95} on the commercial silicon nitride EC-141.}
\label{Fig17}
\end{figure}

Having determined the characteristic size $\ell$ of the degradation region $\Omega_{\ell}(t)$ and the associated material degradation function $\mathcal{D}(\mathfrak{g})$, we are now in a position to deploy the Griffith criticality condition (\ref{Griffith-General-2D}) to compare with the experiments of \cite{Ritchie95}. Figure \ref{Fig17} provides plots of the comparison for the rate ${\rm d}a/{\rm d}N$ of crack growth as a function of the range of stress intensity factors $\Delta K_{\texttt{I}}$. A quick glance at the results suffices to recognize that the theoretical predictions are in good agreement with the experiments for all three load ratios $R=0.1, 0.5, 0.7$. 

%
%

\subsection{Comparisons with the experiments of \cite{Wang91} on mortar}

\cite{Wang91} carried out four-point bending tests on notched beams of length $L=406$ mm, height $H=102$ mm, thickness $B=76$ mm, and initial crack size $A\geq 15$ mm that were made of a mortar with 1 part water, 2 parts cement, and 3.5 parts sand by weight. Table 3 lists the values of the elasticity constants and initial critical energy release rate for this type of material \citep{Giaccio98}. \cite{Wang91} also reported that the largest sand particles did not exceed 3 mm in diameter. The specimens were subjected to a sinusoidal pure bending $M(t)$, parameterized in terms of the global stress measure
\begin{equation*}
\sigma(t)=\dfrac{6M(t)}{B H^2}=\dfrac{\sigma_{max}+\sigma_{min}}{2}+\dfrac{\sigma_{max}-\sigma_{min}}{2}\sin\left(2\pi f_c t-\dfrac{\pi}{2}\right)
\end{equation*}
with minimum stress $\sigma_{min}=0.29$ MPa and loading frequency $f_c=5$ cycles/s. While \cite{Wang91} did not explicitly state the maximum stress $\sigma_{max}$ that he applied, a value of about 50\% of the corresponding critical stress for fracture under monotonic loading is consistent with his data. For this initial-boundary-value problem, the stress intensity factor $K_{\texttt{I}}(t)$ is approximately described by the formula \citep{Tada73}
\begin{equation*}
K_{\texttt{I}}(t)=\sigma(t)\sqrt{\pi a(t)}F\left(a(t)\right)
\end{equation*}
with
\begin{equation}\label{Fa-beam}
F\left(a(t)\right)=1.122-1.4\dfrac{a(t)}{H}+7.33\left(\dfrac{a(t)}{H}\right)^2-13.08\left(\dfrac{a(t)}{H}\right)^3+14\left(\dfrac{a(t)}{H}\right)^4,
\end{equation}
where $a(t)$ is the current size of the crack at time $t$. The main experimental results presented by \cite{Wang91} correspond to the plots in Figs. 4.3.5 and 4.3.8 of his work for the size of the crack $a(t)$, as a function of the number $N$ of loading cycles, in specimens with two different initial crack sizes, $A=22.79$ mm and $32.58$ mm. The values for the maximum stress $\sigma_{max}$ applied in these tests are estimated to be  $\sigma_{max}=2.40$ MPa and $\sigma_{max}=1.94$ MPa, respectively.

%
\begin{figure}[t!]
  \subfigure[]{
   \begin{minipage}[]{0.48\linewidth}
   \centering \includegraphics[width=\linewidth]{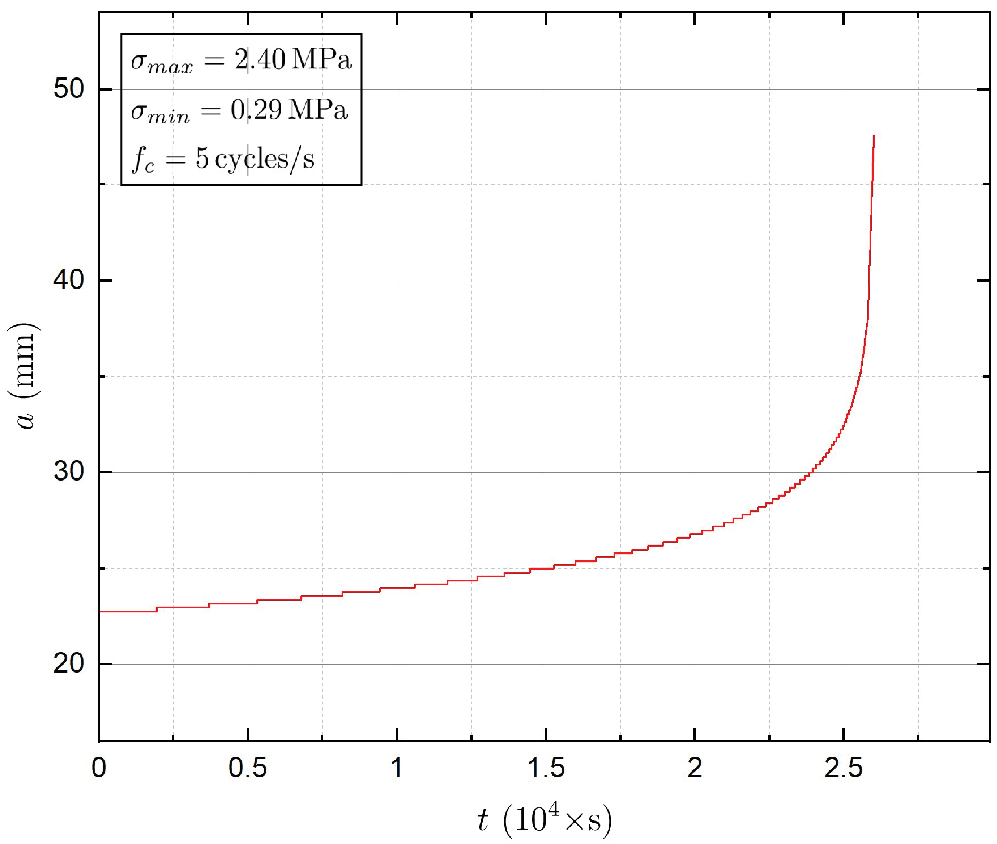}
   \end{minipage}}
  \subfigure[]{
   \begin{minipage}[]{0.48\linewidth}
   \centering \includegraphics[width=\linewidth]{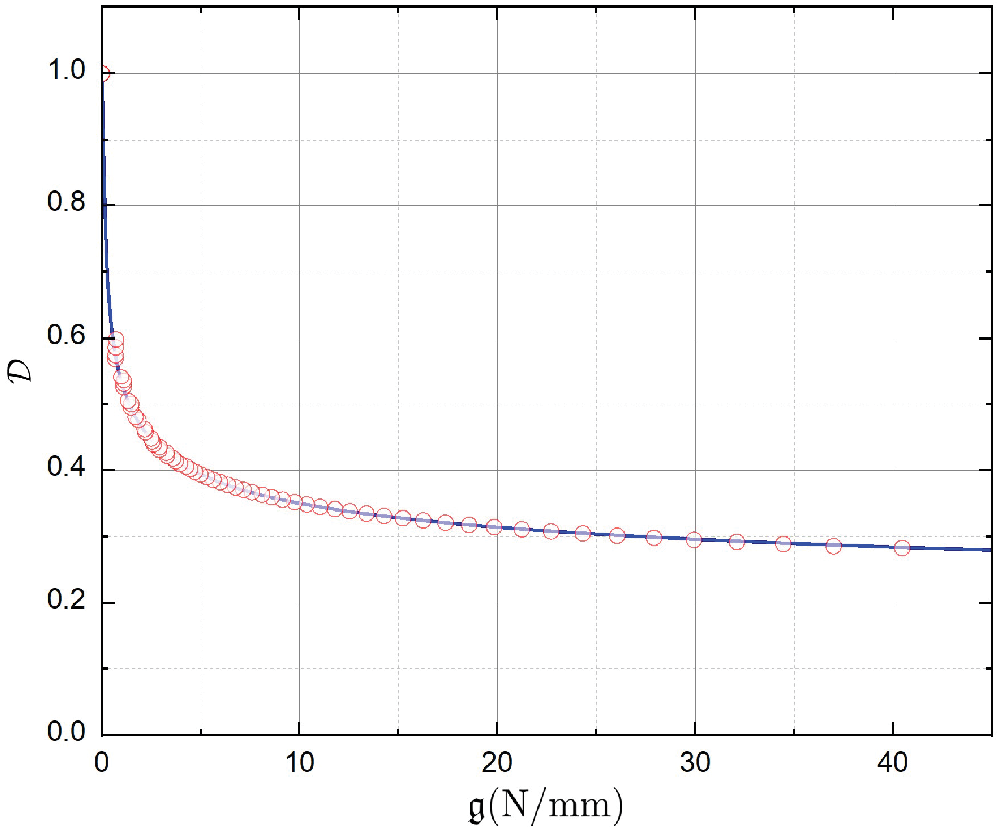}
   \end{minipage}}\par\centering
   \caption{(a) The size $a(t)$ of the crack as a function of time from the four-point bending test of \cite{Wang91} on a notched beam, with initial crack size $A=22.79$ mm, that is made of mortar. The result corresponds to the choice of $\ell=0.2$ mm for the characteristic size of the degradation region $\Omega_{\ell}(t)$. (b) Plot (circles) of the corresponding degradation function $\mathcal{D}(\mathfrak{g})$. For direct comparison, the formula (\ref{D-constitutive-formula}), with the material constants listed in Table \ref{Table6}, is also plotted (solid line).}
   \label{Fig18}
\end{figure}
%

Now, to compare the predictions generated by the Griffith criticality condition (\ref{Griffith-General-2D}) with the experiments of \cite{Wang91}, we again must first determine an appropriate value for the characteristic size $\ell$ of the degradation region $\Omega_{\ell}(t)$ alongside the associated material degradation function $\mathcal{D}(\mathfrak{g})$ for the mortar that he tested. In view of the size of the underlying sand, the largest constituent in the mortar, we set $\ell=0.2$ mm. Given this value, we proceed by fitting one of the crack growth results reported by \cite{Wang91}, say that for initial crack size $A=22.79$ mm. Figure \ref{Fig18}(a) presents a plot of that fit. Then, making use of this data for $a(t)$ and the internal memory variable 
\begin{equation*}%
\mathfrak{g}(t)=\left\{\begin{array}{ll} 
0, & t_j=t \vspace{0.2cm}\\
\dfrac{k_0 \pi a(t_j) F^2(a(t_j))}{E}\left(\sigma^2_{max}-\sigma^2_{min}\right)\left(\left\lfloor f_c t\right\rfloor-\left\lfloor f_c t_j \right\rfloor\right) + c_s(t)-c_s(t_j), & t_j< t<t_{j+1} \end{array}\right. 
\end{equation*}%
pertinent to this problem, where the function $F(a(t))$ is given by (\ref{Fa-beam}) and $c_s(t)$ is a correction term similar to that in (\ref{gt-RCT-Overload}), not spelled out here for conciseness, direct use of the relations 
\begin{equation*}
0\leq \mathcal{D}\left(\mathfrak{g}\right)\leq 1,\quad \mathcal{D}\left(0\right)=1,\quad \displaystyle\lim_{t\nearrow t_{j+1}}\mathcal{D}\left(\mathfrak{g}(t)\right)=\dfrac{\left(\sigma(t_{j+1})\sqrt{\pi a(t_j)}F\left(a(t_j)\right)\right)^2}{G_c E}
\end{equation*}%
allows to construct the material degradation function $\mathcal{D}(\mathfrak{g})$. Figure \ref{Fig18}(b) plots the resulting material degradation function (circles), together with its description (solid line) by the formula  (\ref{D-constitutive-formula}) with the material constants listed in Table \ref{Table6}.

\begin{table}[H]\centering
\caption{Elasticity constants and initial critical energy release rate for mortar.}
\begin{tabular}{cc|c}
\toprule
$E$ (GPa)& $\nu$  & $G_c$ (N/m) \\
\midrule
$30$ & $0.15$ & $55$ \\
\bottomrule
\end{tabular} \label{Table5}
\end{table}
\begin{table}[H]\centering
\caption{The characteristic size $\ell$ of the degradation region $\Omega_{\ell}(t)$ and the associated constants in the material degradation function (\ref{D-constitutive-formula}) for mortar.}
\begin{tabular}{l|ccccc}
\toprule
$\ell$ (mm) & $\dfrac{G_{\infty}}{G_c}$ & $\mathfrak{g}_{th}$ (N/m) & $K$ (m/N)$^{\alpha}$ & $\alpha$ & $\beta$\\
\midrule
$0.2$  & $0.14$ & $55.9$ & $17$ & $0.8242$ & $0.5532$ \\
\bottomrule
\end{tabular} \label{Table6}
\end{table}
%

%
\begin{figure}[H]
\centering \includegraphics[scale=0.5]{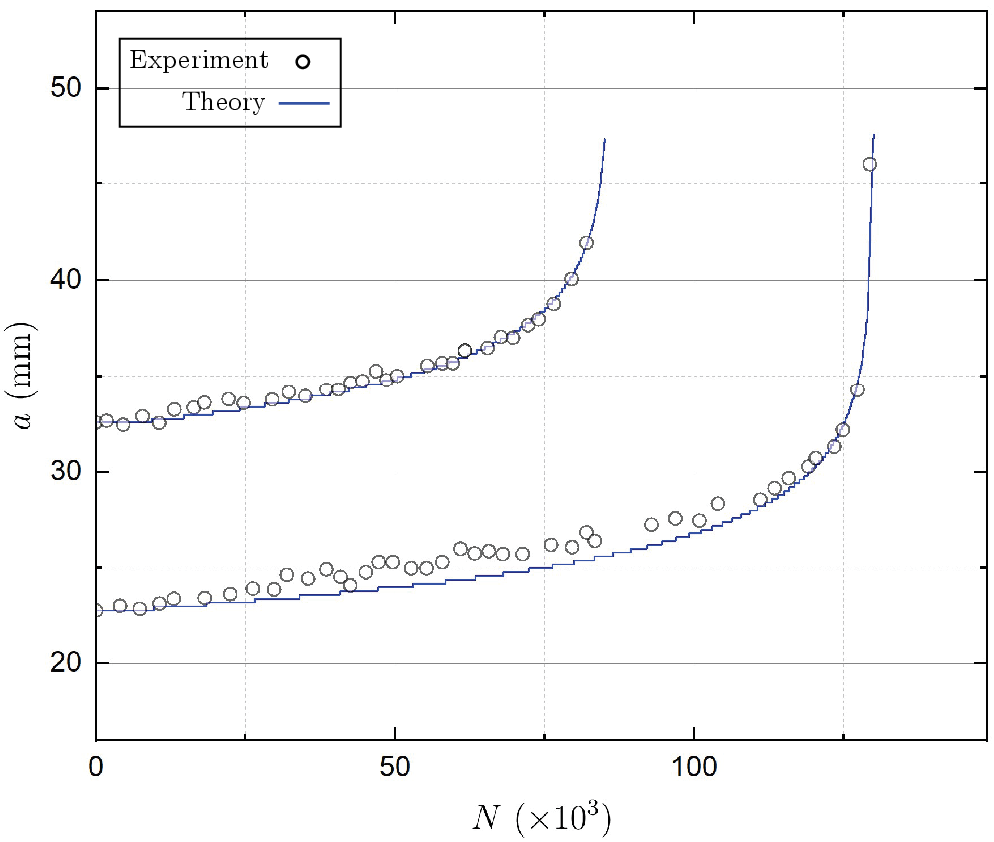}
\caption{Comparison between the predictions (solid lines) generated by the Griffith criticality condition (\ref{Griffith-General-2D}) and the experiments (circles) of \cite{Wang91} on mortar.}
\label{Fig19}
\end{figure}
%

At this point, having determined the characteristic size $\ell$ of the degradation region $\Omega_{\ell}(t)$ and the associated material degradation function $\mathcal{D}(\mathfrak{g})$, we can deploy the Griffith criticality condition (\ref{Griffith-General-2D}) to compare with the experiments of \cite{Wang91}. Figure \ref{Fig19} provides plots of the comparison for the size $a(t)$ of the crack as a function of the number $N$ of loading cycles for the beam with the initial crack size $A=32.58$, as well as for the beam with $A=22.79$ mm used to determine the degradation function. Once again, the agreement between the theoretical predictions and the experimental results is evident.

\subsection{Comparisons with the experiments of \cite{Clark90} on a commercial PMMA}

Finally, \cite{Clark90} performed disk-shaped compact tension tests (see Fig. \ref{Fig8}) on specimens of length $L=84$ mm and thickness $B=6.38$ mm that were made of a commercial PMMA (polymethyl methacrylate) manufactured by Rohm and Haas. No material constants were reported. Here, as listed in Table \ref{Table7}, we make use of typical values for PMMA found in other works in the literature \citep{Dunn97}. The specimens were subjected to the sinusoidal loading (\ref{eq_sin_load-CTT}) with load ratios $R=0.1, 0.65, 0.8$,  no overload $\sigma^\star_{max}=\sigma_{max}$, and loading frequency $f_c=20$ cycles/s. \cite{Clark90} did not report the precise initial size $A$ of the crack in the specimens or the precise value of the maximum stress $\sigma_{max}$ that they applied. A straightforward analysis, however, indicates that the choices $A=25$ mm and $\sigma_{max}=0.3$ MPa are consistent with all their data. We therefore consider that $A=25$ mm and $\sigma_{max}=0.3$ MPa in their experiments. The main experimental results presented by \cite{Clark90} correspond to the three Paris curves that they plotted in Fig. 5(a) of their work for each of the three load ratios $R=0.1, 0.65, 0.8$. 
\begin{table}[H]\centering
\caption{Elasticity constants and initial critical energy release rate for PMMA.}
\begin{tabular}{cc|c}
\toprule
$E$ (GPa)& $\nu$  & $G_c$ (N/m) \\
\midrule
$2.3$ & $0.36$ & $394$ \\
\bottomrule
\end{tabular} \label{Table7}
\end{table}
\begin{table}[H]\centering
\caption{The characteristic size $\ell$ of the degradation region $\Omega_{\ell}(t)$ and the associated constants in the material degradation function (\ref{D-constitutive-formula}) for PMMA.}
\begin{tabular}{l|ccccc}
\toprule
$\ell$ ($\mu$m) & $\dfrac{G_{\infty}}{G_c}$ & $\mathfrak{g}_{th}$ (N/m) & $K$ (m/N)$^{\alpha}$ & $\alpha$ & $\beta$\\
\midrule
$5$  & $0$ & $64$ & $50373$ & $1.1481$ & $0.8833$ \\
\bottomrule
\end{tabular} \label{Table8}
\end{table}

\begin{figure}[H]
\centering \includegraphics[scale=0.5]{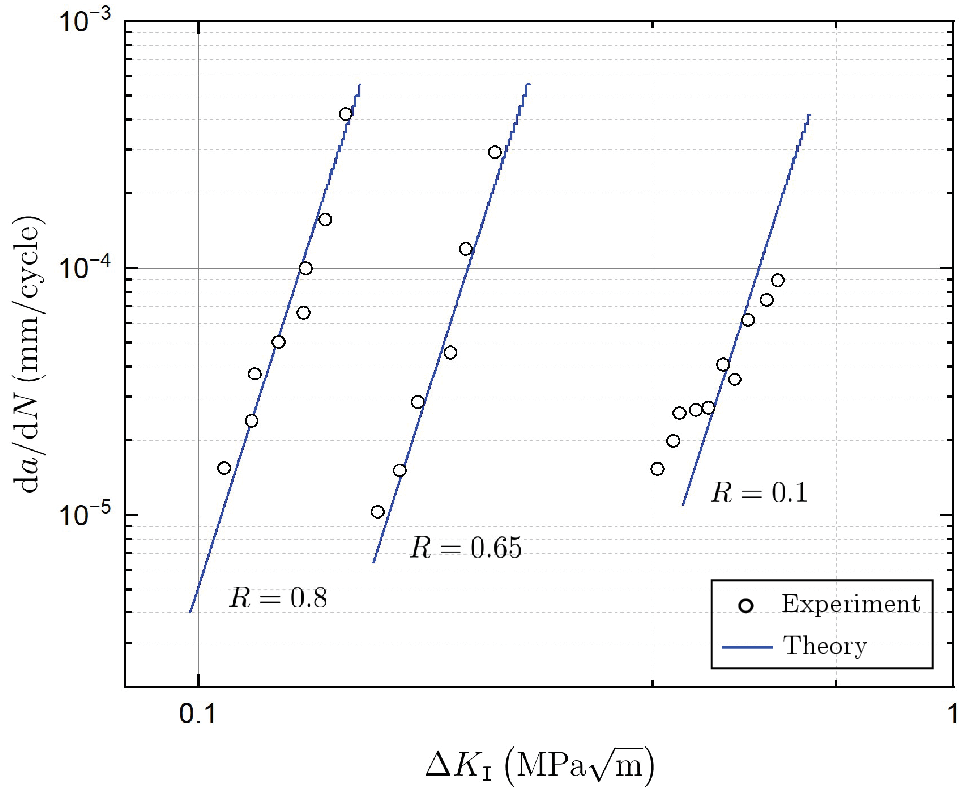}
\caption{Comparison between the predictions (solid lines) generated by the Griffith criticality condition (\ref{Griffith-General-2D}) and the experiments (circles) of \cite{Clark90} on a commercial PMMA.}
\label{Fig20}
\end{figure}

In order to compare the predictions generated by the Griffith criticality condition (\ref{Griffith-General-2D}) with the experiments of \cite{Clark90}, we follow the same steps laid out in the two preceding subsections. In short, we begin by setting the characteristic size of the degradation region $\Omega_\ell(t)$ to $\ell=5$ $\mu$m, since the largest heterogeneity in PMMA is expected to be submicron. We then determine the material degradation function $\mathcal{D}(\mathfrak{g})$ from one of the Paris curves provided by \cite{Clark90}. We choose the Paris curve with load ratio $R=0.8$. The resulting material degradation function $\mathcal{D}(\mathfrak{g})$ can be described with the formula (\ref{D-constitutive-formula}) and the material constants listed in Table \ref{Table8}. Having set $\ell=5$ $\mu$m and having determined $\mathcal{D}(\mathfrak{g})$, we can deploy the Griffith criticality condition (\ref{Griffith-General-2D}) to compare with the  experiments of \cite{Clark90}. Figure \ref{Fig20} presents the comparison for the rate ${\rm d}a/{\rm d}N$ of crack growth as a function of the range of stress intensity factors $\Delta K_{\texttt{I}}$ for all three load ratios $R=0.1, 0.65, 0.8$. Much like the observations from Figs. \ref{Fig17} and \ref{Fig19} above, the main observation from  Fig. \ref{Fig20} is that the theoretical predictions are in good agreement with the experimental results. 

%
%

\section{Summary and final comments} \label{Sec: Final Comments}

In this paper, we have formulated a generalization of the classical Griffith energy competition \citep{Griffith21} to describe the growth of large cracks in nominally elastic brittle materials subjected to arbitrary non-monotonic quasistatic loading conditions. The resulting Griffith criticality condition is given by expression (\ref{Griffith-General}) in its complete form, and by expression (\ref{Griffith-General-2D}) for 2D problems and for 3D problems with a uniformly propagating crack front. 

Guided by an abundance of experimental observations amassed over the past seven decades for a variety of nominally elastic brittle materials ranging from rubbers, to ceramics, to rocks, the three-pronged idea behind the proposed formulation is that:
\begin{enumerate}

\item{the critical energy release rate $\mathcal{G}_c$ is a \emph{not} a material constant --- as assumed under monotonic loading --- but rather a material function $\mathcal{G}_c(\bfX,t)$ of both space and time,}

\item{one whose initially uniform value $\mathcal{G}_c(\bfX,0)=G_c$ decreases as the loading progresses, this solely within a small degradation region $\Omega_\ell(t)$  around crack fronts, with the characteristic size $\ell$ of such a region being material specific, and}

\item{whose decreasing value $\mathcal{G}_c(\bfX,t)=\mathcal{D}\left(\mathfrak{g}(t)\right)G_c$ in $\Omega_\ell(t)$ is described by a material degradation function $\mathcal{D}\left(\mathfrak{g}\right)$ in terms of an internal memory variable $\mathfrak{g}(t)$ that keeps track of the history of the strain field around the evolving crack fronts.}
    
\end{enumerate}

Importantly, the material degradation function $\mathcal{D}\left(\mathfrak{g}\right)$ is an intrinsic macroscopic property of the material. Together with the characteristic size $\ell$ of the degradation region $\Omega_\ell(t)$ and the initial critical energy release rate $G_c$, they completely define the critical energy release rate $\mathcal{G}_c(\bfX,t)$. All these three material properties can be measured by standard tests making use of conventional equipment. Indeed, as exemplified in Section \ref{Sec: Validation}, the characteristic size $\ell$ can be estimated from microscopy images of the largest heterogeneity of the material at hand. The initial critical energy release rate $G_c$ can be obtained from a plethora of standardized tests. Finally, as exemplified in Subsections \ref{Sec: Dg-cali}, \ref{Sec: ell sensitivity}, and Section \ref{Sec: Validation}, the material degradation function $\mathcal{D}\left(\mathfrak{g}\right)$ can be expediently obtained from any of the conventional tests used to generate Paris curves under cyclic loading. Its success to describe the various types of materials that we have considered until now suggests that the expression (\ref{D-constitutive-formula}) may be a useful formula for $\mathcal{D}\left(\mathfrak{g}\right)$. 

By construction, for the limiting case when the loading is monotonic, the proposed Griffith formulation reduces to the classical Griffith formulation. For the opposite limiting case of cyclic loading, the formulation is able to describe any Paris-law behavior of the growth of large cracks observed in experiments. This, together with the properties showcased in Subsection \ref{Sec: Formulation Properties} and the comparisons with experiments presented in Section \ref{Sec: Validation}, provides ample motivation to continue the investigation of the proposed Griffith formulation as a plausible principle in continuum mechanics that describes the growth of large cracks in solids. 

An obvious next step is to work out the proposed Griffith formulation in the setting of finite deformations for nonlinear elastic brittle materials at large and to confront it to a larger set of experiments in order to further scrutinize its predictive capabilities.

Another direction of interest, already mentioned at the beginning of Section \ref{Sec: General}, is to make use of the same three-pronged idea within the variational setting of brittle fracture \citep{Francfort98} so as to work out a formulation that also predicts crack path. As a first step to gain insight into this extension of the work, it is obvious that one can readily implement the proposed Griffith criticality condition in the regularized phase-field formulations of the variational theory \citep{Bourdin00,Bourdin08}. 

Yet another direction of interest is to account for the presence of energy dissipation not just by the creation of new surface but also by deformation, such as, for instance, viscous and plastic deformation. The recent work by \cite{SLP23} should prove invaluable here. 

Finally, following in the footstep of \cite{KFLP18,KBFLP20,KKLP24,LDLP24}, another direction of immediate interest is to incorporate the proposed Griffith formulation into a formulation where the strength of the material is also taken into account so as to work out a complete theory of fracture nucleation and propagation in solids subjected to arbitrary non-monotonic quasistatic loading conditions.

\section*{Acknowledgements}

\noindent This work was funded by the National Science Foundation through the collaborative Grants CMMI--2132528 and CMMI--2132551. This support is gratefully acknowledged.

\bibliographystyle{elsarticle-harv}
\bibliography{References}

\end{document}